\documentclass[11pt,a4paper]{article}
\usepackage{times}
\usepackage{a4wide}
\usepackage{amsfonts}
\usepackage{amssymb}
\usepackage{amsmath}
\usepackage{booktabs} 
\usepackage{etoolbox} 
\AtBeginEnvironment{eqnarray}{\setlength{\arraycolsep}{2pt}}
\usepackage{ifpdf}
\ifpdf
\usepackage[pdftex,unicode,implicit]{hyperref}
\hypersetup{%
  pdftitle    = {Black hole solutions of N=2, d=4 supergravity with a 
quantum correction, in the H-FGK formalism},
  pdfkeywords = {supersymmetry, supergravity, special Kahler geometry, back
    holes, BPS solutions},
  pdfauthor   = {Pietro Galli, Tomas Ortin, Jan Perz and Carlos S. Shahbazi},
  plainpages  = true,
  colorlinks  = true,
  citecolor   = blue,
  urlcolor    = red,
  linkcolor   = black
}

\else
  \usepackage[dvips]{graphicx}
  \usepackage[unicode,implicit]{hyperref}

\fi
\makeatletter
\@addtoreset{equation}{section}
\makeatother

\begin{document}

\begin{flushright}
\small
IFIC/12-75\\
IFT-UAM/CSIC-11-93\\
DFPD-2012-TH-17\\
November 30\textsuperscript{th}, 2012\\
\normalsize
\end{flushright}

\vspace{2cm}

\begin{center}

{\Large {\bf Black-hole solutions of $N=2$, $d=4$ supergravity\\
with a quantum correction, in the H-FGK formalism\par}}

\vspace{2cm}

\renewcommand{\thefootnote}{\alph{footnote}}
{\sl\large
Pietro Galli,$\!^{\dagger}$%
\footnote{E-mail: {\tt Pietro.Galli [at] ific.uv.es}}
Tom\'{a}s Ort\'{\i}n,$\!^{\diamond}$%
\footnote{E-mail: {\tt Tomas.Ortin [at] csic.es}}
Jan Perz$^{\ast}$%
\footnote{E-mail: {\tt Jan.Perz [at] pd.infn.it}}
and Carlos S.~Shahbazi$^{\diamond}$%
\footnote{E-mail: {\tt Carlos.Shabazi [at] uam.es}}}
\setcounter{footnote}{0}
\renewcommand{\thefootnote}{\arabic{footnote}}

\vspace{1cm}

${}^{\dagger}${\it Departament de F\'{\i}sica Te\`orica and IFIC (CSIC-UVEG),
Universitat de Val\`encia,\\
C/ Dr.~Moliner, 50, 46100 Burjassot (Val\`encia), Spain}\\

\vspace{.4cm}

${}^{\diamond}${\it Instituto de F\'{\i}sica Te\'orica UAM/CSIC\\
C/ Nicol\'as Cabrera, 13--15, 28049 Madrid, Spain}\\

\vspace{.4cm}

${}^{\ast}${\it Istituto Nazionale di Fisica Nucleare, Sezione di Padova\\
Via Marzolo, 8, 35131 Padova, Italy}\\

\vspace{2cm}


{\bf Abstract}

\end{center}

\begin{quotation}\small
  We apply the H-FGK formalism to the study of some properties of the
  general class of black holes in $N=2$ supergravity in four
  dimensions that correspond to the harmonic and hyperbolic ans\"atze
  and obtain explicit extremal and non-extremal solutions for the
  $t^3$ model with and without a quantum correction. Not all solutions
  of the corrected model (quantum black holes), including in
  particular a solution with a single $q_{1}$ charge, have a regular
  classical limit.
\end{quotation}

\newpage
\pagestyle{plain}

\tableofcontents


\section{Introduction}

In \cite{Mohaupt:2011aa,Meessen:2011aa} a new formalism for constructing
single-center, static, spherically-symmetric black-hole solutions of $N=2,d=4$
supergravity coupled to vector multiplets was proposed.\footnote{An analogous
  formalism exists for $N=2,d=5$ supergravity theories
  \cite{Mohaupt:2009iq,Mohaupt:2010fk,Meessen:2011aa,Mohaupt:2012tu} and can
  be extended to black-string solutions as well
  \cite{deAntonioMartin:2012bi,Meessen:2012su}.} It is based on rewriting the
effective FGK action \cite{Ferrara:1997tw} in terms of a set of functions
(``$H$-variables'') of the original dynamical fields, chosen in such a way
that they are real and transform linearly under duality. The appropriate
choice, which significantly simplifies the equations of motion, can be made
with the same algorithm for all supergravity prepotentials and for both
extremal and non-extremal black holes. Substituting an ansatz for the
$H$-variables (in \cite{Galli:2011fq} taken to be harmonic and hyperbolic for,
respectively, extremal and non-extremal solutions\footnote{The same ansatz has
  been exploited also in five dimensions
  \cite{Mohaupt:2009iq,Mohaupt:2010fk,Meessen:2011bd,Meessen:2011aa}.})
transforms the equations of motion into a system of ordinary equations on the
parameters of the ansatz in many examples.

This new formalism should simplify considerably the construction of new
black-hole solutions and their systematic study, as it has been shown in the
$N=2,d=5$ case \cite{Mohaupt:2009iq,Mohaupt:2010fk,Meessen:2012su}. So far,
the construction of black-hole solutions demanded the use of a specific ansatz
for each type of solution which had to be plugged into the equations of motion
and checked in detail with considerable effort and, in general, with
meaningful loss of generality, although, eventually, very general ans\"atze
were proposed. The supersymmetric solutions of ungauged $N=2,d=4$ supergravity
coupled to vector supermultiplets, which are the only theories we are going to
study and discuss here, were constructed in this way in a long series of
papers
\cite{Ferrara:1995ih,Strominger:1996kf,Behrndt:1996jn,Sabra:1997kq,Sabra:1997dh,Behrndt:1997ny}. The
effect of the inclusion of $R^{2}$ corrections was studied in
ref.~\cite{LopesCardoso:2000qm}. The outcome of all this work was a very
general recipe that allows the systematic construction of supersymmetric
black-hole solutions from harmonic functions. The same general class of
solutions was eventually shown to contain regular stationary multicenter black
holes \cite{Denef:2000nb,Bates:2003vx}. The complete generality of the
construction has been proven by the use of supersymmetry methods in
\cite{Meessen:2006tu}. For extremal non-supersymmetric black holes
\cite{Khuri:1995xq} no completely general construction procedure is known,
although some general solutions of families of theories have been found such
as for the almost-BPS ones \cite{Ortin:1996bz,Goldstein:2008fq}, those of the
cubic models \cite{Bossard:2012xs} which originate from more particular
examples \cite{LopesCardoso:2007ky,Gimon:2009gk,Bena:2009ev,Galli:2010mg} and the
interacting non-BPS solutions of ref.~\cite{Bossard:2011kz}. In the
non-extremal case, the situation is much worse: only a few examples of general
non-extremal black-hole solutions are known \cite{Galli:2011fq}. The H-FGK
formalism can improve this situation.

Our goals in this paper are similar to those of ref.~\cite{Meessen:2012su} in
the 5-dimensional context: firstly to derive useful model-independent
relationships between the quantities appearing in the H-FGK formalism and the
physical characteristics of the solutions, in sec.~\ref{sec:d4}, and secondly
to use them in sec.~\ref{sec:Example} for finding explicit examples of black
holes in the $t^{3}$ model with a quadratic correction to the prepotential,
whose string-theoretical origin we recall in appendix \ref{sec:CYC}. We
restrict ourselves to solutions described by harmonic and hyperbolic functions
(for the discussion of generality of these ans\"atze see
refs.~\cite{kn:GGP,kn:GMO}). Sec.~\ref{sec:conclusions} contains our
conclusions.


\section{The H-FGK formalism for $N=2$, $d=4$ supergravity}
\label{sec:d4}

In this section we briefly review the H-FGK formalism for theories of
$N=2$, $d=4$ supergravity coupled to $n$ vector multiplets, following
\cite{Meessen:2011aa}, whose conventions we use.

As shown in \cite{Mohaupt:2011aa,Meessen:2011aa}, searching for
single-center, static, spherically symmetric black-hole solutions of
an $N=2$, $d=4$ supergravity coupled to $n$ vector multiplets (and,
correspondingly, including $n$ complex scalars $Z^{i}$ and $n+1$
Abelian vector fields $A^{\Lambda}{}_{\mu}$) with electric
($q_{\Lambda}$) and magnetic ($p^{\Lambda}$) charges described by the
$2(n+1)$-dimensional symplectic vector $(\mathcal{Q}^{M}) \equiv
(p^{\Lambda}, q_{\Lambda})^{\mathrm{T}}$ is equivalent to solving the
following equations of motion for $2(n+1)$ dynamical variables that we
denote by $H^{M}(\tau)$ and identify below with a certain combination of
physical fields:
\begin{eqnarray}
\left(
\partial_{M}\partial_{N}\log{\mathsf{W}}
-2\frac{H_{M}H_{N}}{\mathsf{W}^{2}}
\right)
\ddot{H}^{N}
+
\tfrac{1}{2}\partial_{M}\partial_{N}\partial_{P}\log \mathsf{W}\,
\left(\dot{H}^{N}\dot{H}^{P} -\tfrac{1}{2}\mathcal{Q}^{N}\mathcal{Q}^{P}
\right)
& & \nonumber \\
& & \nonumber \\
-4\dot{H}_{M}\frac{\dot{H}^{N}H_{N}}{\mathsf{W}^{2}}
+8H_{M}\frac{\dot{H}^{P}\tilde{H}_{P}\, \dot{H}^{N}H_{N}}{\mathsf{W}^{3}}
+2 \mathcal{Q}_{M} \frac{H^{N}\mathcal{Q}_{N}}{\mathsf{W}^{2}}
& & \nonumber \\
& & \nonumber \\
-4\tilde{H}_{M}\frac{(H^{N}\dot{H}_{N})^{2}}{\mathsf{W}^{3}}
-4\tilde{H}_{M}\frac{(H^{N}\mathcal{Q}_{N})^{2}}{\mathsf{W}^{3}}
& = &
0\, ,
\label{eq:equationsofmotion}
\\
& & \nonumber \\
-\tfrac{1}{2}\partial_{M}\partial_{N}\log\mathsf{W}
\left(\dot{H}^{M}\dot{H}^{N}-\tfrac{1}{2}\mathcal{Q}^{M}\mathcal{Q}^{N} \right)
+\left(\frac{\dot{H}^{M}H_{M}}{ \mathsf{W}}\right)^{2}
-\left(\frac{\mathcal{Q}^{M}H_{M}}{ \mathsf{W}}\right)^{2}
& = &
r_{0}^{2}\, .
\label{eq:hamiltonianconstraint}
\end{eqnarray}
\noindent
In these equations $r_{0}$ is the non-extremality parameter, we use
the symplectic form $\left(\Omega_{MN} \right) \equiv
\bigl(\begin{smallmatrix} 0 & \mathbb{I}\\ -\mathbb{I} & 0
\end{smallmatrix} \bigr)$ and $\Omega^{MN}=\Omega_{MN}$ to lower and
raise the symplectic indices according to the convention
\begin{equation}
H_{M} = \Omega_{MN}H^{N}\, ,
\qquad
H^{M} = H_{N}\Omega^{NM}\, ,
\end{equation}
\noindent
and $\mathsf{W}(H)$ is the \textit{Hesse potential}.\footnote{For a
  historical perspective on the real formulation of special K\"ahler
  geometry and the Hesse potential see
  e.g.~\cite{Mohaupt:2011aa,Cardoso:2012nh}.} For a theory defined by
the covariantly holomorphic symplectic section $\mathcal{V}^{M}$, the
Hesse potential can be found as follows: introducing a complex
variable $X$ with the same K\"ahler weight as $\mathcal{V}^{M}$, we
can define the K\"ahler-neutral real symplectic vectors
\begin{equation}
\label{eq:RandIdef}
\mathcal{R}^{M}  = \Re\mathfrak{e}\, \mathcal{V}^{M}/X\, ,
\qquad
\mathcal{I}^{M} =\Im\mathfrak{m}\,  \mathcal{V}^{M}/X\, .
\end{equation}
\noindent
The components $\mathcal{R}^{M}$ can be expressed in terms of the
$\mathcal{I}^{M}$, to which process we refer later as solving
Freudenthal duality equations.\footnote{In earlier papers sometimes
  called ``stabilization equations''.} Then, the Hesse potential, as
a function of the components $\mathcal{I}^{M}$ is given by
\begin{equation}
\label{eq:Wdef}
\mathsf{W}(\mathcal{I})
\equiv
\langle\, \mathcal{R}(\mathcal{I})\mid \mathcal{I} \,  \rangle
\equiv
\mathcal{R}_{M}(\mathcal{I})\mathcal{I}^{M}\, ,
\end{equation}
\noindent
and identifying $\mathcal{I}^{M}=H^{M}$ we get $\mathsf{W}(H)$. We
can use $\mathcal{R}^{M}$ to define dual variables:
\begin{equation}
\tilde{H}^{M}(H) \equiv \mathcal{R}^{M}(H)\, .
\end{equation}

Given a solution $H^{M}(\tau)$ of the equations
(\ref{eq:equationsofmotion}) and (\ref{eq:hamiltonianconstraint}), the
warp factor $e^{2U}$ of the spacetime metric
\begin{equation}
\label{eq:generalbhmetric}
ds^{2} =
e^{2U} dt^{2} - e^{-2U}\left(\frac{r_{0}^{4}}{\sinh^{4} r_{0}\tau}d\tau^{2}
+
\frac{r_{0} ^{2}}{\sinh^{2}r_{0}\tau}d\Omega^{2}_{(2)}\right) ,
\end{equation}
\noindent
takes the form
\begin{equation}
e^{-2U}= \mathsf{W}(H)
\end{equation}
\noindent
and the scalar fields are given by
\begin{equation}
Z^{i} = \frac{\tilde{H}^{i}+iH^{i}}{\tilde{H}^{0} +iH^{0}}\, .
\end{equation}

The equations of motion (\ref{eq:equationsofmotion}) can be derived from the
effective action
\begin{equation}
\label{eq:effectiveaction2}
-I_{\rm eff}[H]
=
\int d\tau
\left[
\tfrac{1}{2}\partial_{M}\partial_{N}\log\mathsf{W}
\left(\dot{H}^{M}\dot{H}^{N}+\tfrac{1}{2}\mathcal{Q}^{M}\mathcal{Q}^{N} \right)
\noindent
- \left(\frac{\dot{H}^{M}H_{M}}{ \mathsf{W}}\right)^{2}
-\left(\frac{\mathcal{Q}^{M}H_{M}}{ \mathsf{W}}\right)^{2}
\right] .
\end{equation}
\noindent
Then, eq.~(\ref{eq:hamiltonianconstraint}) is nothing but the
Hamiltonian constraint associated with the $\tau$-independence of the
action, with a particular value of the integration constant, which we
cannot change because it is part of the transverse metric ansatz.

If we contract the equations of motion (\ref{eq:equationsofmotion}) with
$H^{P}$ and use the homogeneity properties of the different terms and the
Hamiltonian constraint eq.~(\ref{eq:hamiltonianconstraint}), we find a useful
equation
\begin{equation}
\label{eq:Urewriten}
\tilde{H}_{M}\left(\ddot{H}^{M}
  -r_{0}^{2}H^{M}\right)
+\frac{(\dot{H}^{M}H_{M})^{2}}{\mathsf{W}} = 0\, ,
\end{equation}
\noindent
which corresponds to that of the variable $U$ minus the Hamiltonian constraint
in the standard formulation.\footnote{This equation in the extremal limit
  agrees with the special static case of eq.~(3.31) of
  ref.~\cite{Galli:2010mg}.}

In what follows we shall impose on the variables $H^{M}$ the constraint
\begin{equation}
\label{eq:nonutconstraint}
\dot{H}^{M}H_{M} = 0\, .
\end{equation}
\noindent
In the supersymmetric (hence, extremal) case it has been shown
\cite{Bellorin:2006xr} that this constraint enforces the absence of NUT
charge: a non-zero NUT charge would lead to a non-static metric with
string-like singularities.  Here, this condition is nothing but a possible
simplifying assumption which does not imply non-staticity since staticity has
been assumed in this formalism form the onset. Here we take it as a convenient
ansatz and leave the possibility and implications of violating this constraint
to be studied elsewhere \cite{kn:GGP,kn:GMO}.

The above constraint simplifies eq.~(\ref{eq:Urewriten})
\begin{equation}
\label{eq:Urewriten2}
\tilde{H}_{M}\left(\ddot{H}^{M}
  -r_{0}^{2}H^{M}\right)
= 0\, ,
\end{equation}
\noindent
which can be solved by harmonic (in the extremal $r_{0}=0$ case) or
hyperbolic (in the non-extremal $r_{0}\neq 0$ case) ans\"atze for the
variables $H^{M}$, satisfying
\begin{equation}
\ddot{H}^{M}  -r_{0}^{2}H^{M} =  0 \, .
\end{equation}

These are the ans\"atze that we will use in the rest of the paper,
bearing in mind that they are adapted to the additional constraint
(\ref{eq:nonutconstraint}) that we impose by hand. Taking into account
this constraint, the equations that need to be solved are:
\begin{eqnarray}
\partial_{P}\partial_{M}\log \mathsf{W}\, \ddot{H}^{M}
+\tfrac{1}{2}\partial_{P}\partial_{M}\partial_{N}\log \mathsf{W}\,
\left(\dot{H}^{M}\dot{H}^{N} -\tfrac{1}{2}\mathcal{Q}^{M}\mathcal{Q}^{N}
\right)
+\partial_{P}\!\left(\!\frac{\mathcal{Q}^{M}H_{M}}{\mathsf{W}}\!\right)^{2}
& = &
0\, ,
\label{eq:equationsofmotion2}
\\
& & \nonumber \\
-\tfrac{1}{2}\partial_{M}\partial_{N}\log\mathsf{W}
\left(\dot{H}^{M}\dot{H}^{N}-\tfrac{1}{2}\mathcal{Q}^{M}\mathcal{Q}^{N} \right)
-\left(\!\frac{\mathcal{Q}^{M}H_{M}}{ \mathsf{W}}\!\right)^{2}
& = &
r_{0}^{2}\, ,
\label{eq:hamiltonianconstraint2}
\\
& & \nonumber \\
\dot{H}^{M}H_{M}
& = &
0\, .
\label{eq:noNUTconstraint}
\end{eqnarray}
It is also useful to have the expression of the black-hole potential as a
zeroth-degree homogeneous function of the variables $H^{M}$:
\begin{equation}
-V_{\rm bh}(H,\mathcal{Q})
=
-\tfrac{1}{4} \mathsf{W}
\left(
\partial_{M}\partial_{N}\log{\mathsf{W}} -4 \mathsf{W}^{-2}H_{M}H_{N}
\right)
 \mathcal{Q}^{M}
 \mathcal{Q}^{N}\, .
\end{equation}


\subsection{Extremal black holes}
\label{sec:d4extremal}

As explained above, for extremal black holes we take $H^{M}(\tau)$ to be
harmonic in Euclidean $\mathbb{R}^{3}$, i.e.~linear in
$\tau$:\footnote{\label{foot:laotra}Known non-supersymmetric extremal solutions
  that do not conform to this ansatz do not satisfy constraint
  (\ref{eq:nonutconstraint}) either \cite{Galli:2010mg,kn:GGP}. On the other
  hand, the representation of a solution in terms of the $H^{M}$ may not be
  unique and the harmonicity or the fact that the constraint
  eq.~(\ref{eq:noNUTconstraint}) is satisfied may not always be a
  characteristic feature of a solution \cite{kn:GMO}.}
\begin{equation}
\label{eq:harmonicfunctions}
H^{M}= A^{M} -\tfrac{1}{\sqrt{2}}B^{M}\tau\, ,
\end{equation}
\noindent
where $A^{M}$ and $B^{M}$ are integration constants to be determined as
functions of the independent physical constants (namely, the charges
$\mathcal{Q}^{M}$ and the values of the scalars at spatial infinity
$Z^{i}_{\infty}$) by using the equations of motion
(\ref{eq:equationsofmotion2})--(\ref{eq:noNUTconstraint}) and the asymptotic
conditions.

The equations of motion for the above ansatz can be written in a simple and
suggestive form\footnote{\label{foot:esa}It is worth stressing that, even
  though the first equation is the derivative of the second with respect to
  $H^{P}$, solving the second for some functions $H^{M}$ does not imply having
  solved the first. Only if we find a $B^{M}$ such that the second equation is
  satisfied identically for any $H^{M}$ will the first equation be satisfied as
  well. The number of $B^{M}$ with this property and their value depend on the
  particular theory under consideration, but their existence is quite a general
  phenomenon.}
\begin{eqnarray}
\partial_{P}\left[V_{\rm bh}(H,\mathcal{Q}) - V_{\rm bh}(H,B)\right]
& = &
0\, ,
\label{eq:dV=dV}
\\
& & \nonumber \\
V_{\rm bh}(H,\mathcal{Q})- V_{\rm bh}(H,B)
& = &
0\, ,
\label{eq:V=V}
\\
& & \nonumber \\
A^{M}B_{M}
& = &
0\, .
\label{eq:AB=0}
\end{eqnarray}
\noindent
Observe that the first two equations are automatically solved for
$B^{M}=\mathcal{Q}^{M}$, which corresponds to the supersymmetric case. The
third equation then takes the form $A^{M}\mathcal{Q}_{M}$ and still has to be
solved, which can be done generically \cite{Denef:2000nb,Bates:2003vx} as we
are going to show.

Furthermore, observe that the Hamiltonian constraint (\ref{eq:V=V}) is
equivalent to the requirement that the black-hole potential
\textit{evaluated on the solutions} has the same form in terms of the
\textit{fake central charge} which we can define for any symplectic
(\textit{fake} or not \textit{fake}) charge vector $B^{M}$ by
\begin{equation}
\label{eq:fakecentralchargedef}
\tilde{\mathcal{Z}}(Z,Z^{*},B) \equiv \langle\, \mathcal{V}\mid B\, \rangle
\end{equation}
\noindent
as in terms of the actual central charge $\mathcal{Z}(Z,Z^{*},Q)
\equiv \langle\, \mathcal{V}\mid Q\, \rangle =
\tilde{\mathcal{Z}}(Z,Z^{*},\mathcal{Q})$, that is
\begin{equation}
\label{eq:VandfakeZ}
-V_{\rm bh}(Z,Z^{*},\mathcal{Q})
=
|\tilde{\cal Z}|^{2}
+\mathcal{G}^{ij^{*}}\mathcal{D}_{i}\tilde{\cal Z}\,\mathcal{D}_{j^{*}}\tilde{\cal Z}^{*}\, .
\end{equation}

The asymptotic conditions take the form
\begin{eqnarray}
\label{eq:asympflat}
\mathsf{W}(A)
& = &
1\, ,
\\
& & \nonumber \\
Z^{i}_{\infty}
& = &
\frac{\tilde{H}^{i}(A)+iA^{i}}{\tilde{H}^{0}(A) +iA^{0}}\, ,
\end{eqnarray}
\noindent
but can always be solved, together with (\ref{eq:AB=0}), as follows: if we
write $X$ as
\begin{equation}
\label{eq:alphadef}
X = \tfrac{1}{\sqrt{2}} e^{U+i\alpha}\, ,
\end{equation}
\noindent
then, from the definition (\ref{eq:RandIdef}) of $\mathcal{I}^{M}$ we get
\begin{equation}
H^{M} = \sqrt{2} e^{-U}\, \Im\mathfrak{m}(e^{-i\alpha} \mathcal{V}^{M})\, ,
\end{equation}
\noindent
and, at spatial infinity $\tau=0$, using asymptotic flatness
(\ref{eq:asympflat})
\begin{equation}
A^{M} = \sqrt{2} \, \Im\mathfrak{m}(e^{-i\alpha_{\infty}} \mathcal{V}^{M}_{\infty})\, .
\end{equation}
\noindent
Now, to determine $\alpha_{\infty}$ we can use (\ref{eq:AB=0}) and the
definition of fake central charge (\ref{eq:fakecentralchargedef}).  Observe
that
\begin{equation}
A_{M}B^{M}
=
\langle\, H\mid B\, \rangle
=
\Im\mathfrak{m} \langle\, \mathcal{V}/X \mid B\,\rangle
=
\Im\mathfrak{m} (\tilde{\mathcal{Z}}/X ) =
\sqrt{2}e^{-U}\Im\mathfrak{m} (e^{-i\alpha} \tilde{\mathcal{Z}} )
=0\, ,
\end{equation}
\noindent
from which one first obtains the relation
\begin{equation}
\label{eq:alpha}
e^{i\alpha}= \pm \tilde{\mathcal{Z}}/|\tilde{\mathcal{Z}}|
\end{equation}
\noindent
and then the general expression for the $A^{M}$ as a function of the
$B^{M}$ and the $Z^{i}_{\infty}$:
\begin{equation}
\label{eq:AMextremal}
A^{M} = \pm \sqrt{2} \,
\Im\mathfrak{m} \left(\frac{\tilde{\mathcal{Z}}^{*}_{\infty}}{|\tilde{\mathcal{Z}}_{\infty}|}
\mathcal{V}^{M}_{\infty}\right) .
\end{equation}
\noindent
The sign of $A^{M}$ should be chosen to make $H^{M}$ finite (and, generically,
the metric non-singular) in the range $\tau\in (-\infty,0)$. The positivity of
the mass is a physical condition that eliminates some singularities of the
metric. As we shall see in eq.~(\ref{eq:massformulad4-1}), this
requirement singles out the upper sign in the above formula.

Having reduced the problem of finding a complete solution to the
determination of the constants $B^{M}$ that must satisfy equations
(\ref{eq:dV=dV}) and (\ref{eq:V=V}) as functions of the physical
parameters $\mathcal{Q}^{M},Z^{i}_{\infty}$, it is useful to analyze
the near-horizon and spatial-infinity limits of these two
equations. The near-horizon limit of (\ref{eq:V=V}) plus the
definition of the fake central charge lead to the following chain of
relations\footnote{In this and other equations, the expression $V_{\rm
    bh}(B,\mathcal{Q})$ stands for the standard black-hole
  potential with the functions $H^{M}(\tau)$ replaced by the constants
  $B^{M}$.}
\begin{equation}
\label{eq:AvsVh4d}
S/\pi
=
\tfrac{1}{2}\mathsf{W}(B)
=
-V_{\rm bh}(B,\mathcal{Q})
=
|\tilde{\cal Z}(B,B)|^{2}\, ,
\end{equation}
\noindent
where $S$ is the Bekenstein--Hawking black hole entropy and $\tilde{\cal
  Z}(B,B)$ is the near-horizon value of the fake central charge. The last of
these relations, together with the condition (\ref{eq:VandfakeZ}) imply that,
on the horizon, the fake central charge reaches an extremum
\begin{equation}
\partial_{i}|\tilde{\cal Z}(Z_{\mathrm{h}},Z^{*}_{\mathrm{h}},B)|=0\, .
\end{equation}

The near-horizon limit of  (\ref{eq:dV=dV}) leads to
\begin{equation}
\label{eq:dVh=0}
\partial_{M}V_{\rm bh}(B,\mathcal{Q})=0\, ,
\end{equation}
\noindent
which says that the $B^{M}$ extremize the value of the black-hole potential on
the horizon. Since the black-hole potential is invariant under a global
rescaling of the $H^{M}$, the solutions (that we generically call
\textit{attractors} $B^{M}$) of these equations are determined up to a global
rescaling, which can be fixed by imposing eq.~(\ref{eq:V=V}).

The $B^{M}$ must transform under the duality group of the theory (embedded in
$Sp(2n+2,\mathbb{R})$) in the same representation as the $H^{M}$, the charges
$\mathcal{Q}^{M}$ and the constants $A^{M}$. In certain cases this poses strong
constraints on the possible solutions, since building from $\mathcal{Q}^{M}$
and $Z^{i}_{\infty}$ an object that transforms in the right representation of
the duality group and has dimensions of length squared may be far from
trivial. A possibility that is always available is the Freudenthal dual defined
in ref.~\cite{Ferrara:2011gv}, generalizing the definition made in
ref.~\cite{Borsten:2009zy}. Freudenthal duality in $N=2,d=4$ theories
can be understood as the transformation from the $H^{M}$ to the
$\tilde{H}_{M}(H)$ variables. The same transformation can be applied to any
symplectic vector, such as the charge vector. Then, in our notation and
conventions, the Freudenthal dual of the charge vector, $\tilde{\mathcal{Q}}_{M}$,
is defined by
\begin{equation}
\tilde{\mathcal{Q}}_{M} =
\frac{1}{2}\frac{\partial \mathsf{W}(\mathcal{Q})}{\partial \mathcal{Q}^{M}}\, .
\end{equation}

It is not difficult to prove that this duality transformation is an
antiinvolution
\begin{equation}
\tilde{\tilde{\mathcal{Q}}}_{M} = -\mathcal{Q}_{M}\, ,
\end{equation}
\noindent
and using eq.~(\ref{eq:Wdef}) to show that
\begin{equation}
\mathsf{W}(\tilde{\mathcal{Q}}) = \mathsf{W}(\mathcal{Q})\, .
\end{equation}

With more effort one can also show that the critical points of the black-hole
potential are invariant under Freudenthal duality
\cite{Ferrara:2011gv}. Therefore, as $B^{M}=\mathcal{Q}^{M}$ is always an
attractor (the supersymmetric one),
\begin{equation}
B^{M} = \tilde{\mathcal{Q}}^{M}
\end{equation}
\noindent
will always be another attractor.

Let us now consider the spatial-infinity limit, taking into account the
definition of the mass in these spacetimes and the definition of the fake
central charge
\begin{equation}
\label{eq:massformulad4-1}
M=\dot{U}(0)
= \tfrac{1}{\sqrt{2}} \langle\, \tilde{A} \mid B\,\rangle
= \pm  |\tilde{\cal Z}(A,B)| \, .
\end{equation}
\noindent
As mentioned before, to have a positive mass we must use exclusively the upper
sign in (\ref{eq:alpha}) and (\ref{eq:AMextremal}) and we do so from now
onwards. In the supersymmetric case, when $B^{M}=\mathcal{Q}^{M}$ and the fake
central charge becomes the true one, this is the supersymmetric BPS relation.

The asymptotic limit of (\ref{eq:V=V}) plus (\ref{eq:VandfakeZ}) and the above
relation give
\begin{equation}
M^{2} + \left[\mathcal{G}^{ij^{*}}\mathcal{D}_{i}\tilde{\cal Z}\,
\mathcal{D}_{j^{*}}\tilde{\cal Z}^{*}\right]_{\infty}
+V_{\rm bh\, \infty} =0\, ,
\end{equation}
\noindent
which, when compared with the general BPS bound \cite{Ferrara:1997tw}, leads to
the identification of the scalar charges $\Sigma_{i}$ with the values of the
covariant derivatives of the fake central charges at spatial infinity
\begin{equation}
\Sigma_{i} = \left. \mathcal{D}_{i}\tilde{\cal Z}  \right|_{\infty}\, .
\end{equation}


\subsubsection{First-order flow equations}
\label{First_order_ext}

First-order flow equations for extremal BPS and non-BPS black holes can be
easily found following \cite{Ortin:2011vm} but using the generic harmonic
functions (\ref{eq:harmonicfunctions}): let us consider the K\"ahler-covariant
derivative of the inverse of the auxiliary function
\begin{equation}
  \begin{array}{rcl}
\mathcal{D}X^{-1}
& = &
i \langle\, \mathcal{V} \mid \mathcal{V}^{*}\rangle \mathcal{D}X^{-1}
=
i \langle\, \mathcal{D}(\mathcal{V}/X) \mid \mathcal{V}^{*}\rangle
=
i \langle\, d(\mathcal{V}/X) \mid \mathcal{V}^{*}\rangle
\\
& & \\
& = &
i \langle\, d(\mathcal{V}/X) - d(\mathcal{V}/X)^{*} \mid
\mathcal{V}^{*}\rangle
=
-2 \langle\, d H \mid \mathcal{V}^{*}\rangle
\\
& & \\
& = &
-\sqrt{2}\, \tilde{\cal Z}^{*}(Z,Z^{*},B)\, d\tau\, ,
\end{array}
\end{equation}
\noindent
where we have used the normalization of the symplectic section in the first
step, the property $\langle\, \mathcal{D}\mathcal{V} \mid
\mathcal{V}^{*}\rangle =0$ in the second, the K\"ahler-neutrality of
$\mathcal{V}/X$ in the third, $\langle\, \mathcal{D}\mathcal{V}^{*} \mid
\mathcal{V}^{*}\rangle = \langle\, \mathcal{V}^{*} \mid \mathcal{V}^{*}\rangle
= 0$ in the fourth, the definition of $\mathcal{I}=H$ in the fifth, and the
ansatz (\ref{eq:harmonicfunctions}) and the definition of the fake central
charge (\ref{eq:fakecentralchargedef}) in the sixth.

From this equation, eqs.~(\ref{eq:alphadef}) and (\ref{eq:alpha}) and
the relation (cf.~eqs.~(3.8), (3.28) in ref.~\cite{Galli:2010mg})
\begin{equation}
 \dot{\alpha} = -\mathcal{Q}_{\star}\, , 
\hspace{.5cm}\mbox{where}\hspace{.5cm}
\mathcal{Q}_{\star} = \tfrac{1}{2i}\dot{Z}^{i}\partial_{i}\mathcal{K} +\mathrm{c.c.} 
\end{equation}
\noindent
is the pullback of the K\"ahler connection 1-form, we find the
standard first-order equation for the metric function $U$:
\begin{equation}\label{1st_order_Uext}
\frac{de^{-U}}{d\tau } = -|\tilde{\cal Z}(Z,Z^{*},B)|\, .
\end{equation}

Let us now consider the differential of the complex scalar fields:
\begin{equation}
  \begin{array}{rcl}
dZ^{i}
& = &
i \mathcal{G}^{ij^{*}}
\langle\, \mathcal{D}_{j^{*}}\mathcal{V}^{*} \mid
\mathcal{D}_{k}\mathcal{V}\, \rangle
dZ^{k}
=
iX \mathcal{G}^{ij^{*}}
\langle\, \mathcal{D}_{j^{*}}\mathcal{V}^{*} \mid
\mathcal{D}_{k}(\mathcal{V}/X)\, \rangle
dZ^{k}
\\
& & \\
& = &
iX \mathcal{G}^{ij^{*}}
\langle\, \mathcal{D}_{j^{*}}\mathcal{V}^{*} \mid
\partial_{k}(\mathcal{V}/X)\, \rangle
dZ^{k}
=
iX \mathcal{G}^{ij^{*}}
\langle\, \mathcal{D}_{j^{*}}\mathcal{V}^{*} \mid
d(\mathcal{V}/X)\, \rangle
\\
& & \\
& = &
iX \mathcal{G}^{ij^{*}}
\langle\, \mathcal{D}_{j^{*}}\mathcal{V}^{*} \mid
d(\mathcal{V}/X)-d(\mathcal{V}/X)^{*}\, \rangle
=
-2X \mathcal{G}^{ij^{*}}
\langle\, \mathcal{D}_{j^{*}}\mathcal{V}^{*} \mid
dH\, \rangle
\\
& & \\
& = &
+\sqrt{2}\, X \mathcal{G}^{ij^{*}}
\langle\, \mathcal{D}_{j^{*}}\mathcal{V}^{*} \mid
B \, \rangle \, d\tau
=
\sqrt{2}\, X \mathcal{G}^{ij^{*}}
\mathcal{D}_{j^{*}}\tilde{\cal  Z}^{*}(Z,Z^{*},B)\,
 d\tau\, ,
\end{array}
\end{equation}
\noindent
where we have used the same properties as before. To put this expression in a
more conventional form we can use the covariant holomorphicity of $\tilde{\cal
  Z}$ writing
\begin{equation}
\mathcal{D}_{j^{*}}\tilde{\cal  Z}^{*}
=
\mathcal{D}_{j^{*}}\frac{|\tilde{\cal  Z}|^{2}}{\tilde{\cal  Z}}
=
\frac{2 |\tilde{\cal  Z}| \partial_{j^{*}}|\tilde{\cal  Z}|}{\tilde{\cal  Z}}
=2 e^{-i\alpha}\partial_{j^{*}}|\tilde{\cal  Z}|\, ,
\end{equation}
\noindent
and plugging this result in the expression above:
\begin{equation}\label{1st_order_Zext}
\frac{dZ^{i}}{d\tau}
=
2e^{U} \mathcal{G}^{ij^{*}}
\partial_{j^{*}}|\tilde{\cal  Z}|\, .
\end{equation}

It is easy to check that these first order equations imply the second-order
equations of motion
\begin{eqnarray}
\label{2nd_eq_U}\ddot{U}
+e^{2U}V_{\rm bh}(Z,Z^{*},B)
& = & 0\, ,\\
& & \nonumber \\
\label{2nd_eq_Z}\ddot{Z}^{i}
+\Gamma_{jk}{}^{i} \dot{Z}^{j} \dot{Z}^{k}
+e^{2U}\partial^{i}V_{\rm bh}(Z,Z^{*},B)
& = & 0\, ,
\end{eqnarray}
\noindent
with $\Gamma_{jk}{}^{i} =
\mathcal{G}^{il^{*}}\partial_{j}\mathcal{G}_{k l^{*}}$, which coincide
with the original ones if
\begin{equation}
V_{\rm bh}(Z,Z^{*},B) = V_{\rm bh}(Z,Z^{*},\mathcal{Q})
\end{equation}
\noindent
for any $Z^{i}$ (not just for the solution; see the remark in footnote
\ref{foot:esa}).


\subsection{Non-extremal black holes}
\label{sec:d4ansatz}

Previous experience \cite{Galli:2011fq} (see also \cite{Mohaupt:2011aa} and,
further, \cite{Meessen:2011bd,Meessen:2012su} for 5-dimensional examples)
suggests that a suitable ansatz for the variables $H^{M}$ for non-extremal
black holes of $N=2,d=4$ supergravity, compatible with the constraint
(\ref{eq:nonutconstraint}), is
\begin{equation}
\label{eq:Ansatz}
H^{M}(\tau)
=
A^{M}\cosh(r_{0}\tau) + \frac{B^{M}}{r_{0}}\sinh(r_{0}\tau)\, ,
\end{equation}
\noindent
for some integration constants $A^{M}$ and $B^{M}$ that, as in the extremal
case, have to be determined by solving the equations of motion and by imposing
the standard normalization of the physical fields at spatial infinity.

Using this ansatz, the equations of motion
(\ref{eq:equationsofmotion2})--(\ref{eq:noNUTconstraint}) take the form
\begin{eqnarray}
\tfrac{1}{2}\partial_{P}\partial_{M}\partial_{N}\log \mathsf{W}\,
\left(B^{M}B^{N} -r_{0}^{2}A^{M}A^{N}
\right)
-\partial_{P}\left[V_{\rm bh}(Z,Z^{*},\mathcal{Q})/\mathsf{W} \right]
 & = &
0\, ,
\label{eq:equationsofmotion2nonext}
\\
& & \nonumber \\
-\tfrac{1}{2}\partial_{M}\partial_{N}\log\mathsf{W}
\left(B^{M}B^{N} -r_{0}^{2}A^{M}A^{N}\right)
-V_{\rm bh}(Z,Z^{*},\mathcal{Q})/\mathsf{W}
& = &
0\, ,
\label{eq:hamiltonianconstraint2nonext}
\\
& & \nonumber \\
A^{M} B_{M}
& = &
0\, ,
\label{eq:noNUTconstraintnonext}
\end{eqnarray}
\noindent
where we have used the third equation and the homogeneity properties of the
Hesse potential $\mathsf{W}$ in order to simplify the first two.

In the non-extremal case we can define several fake central charges:
\begin{equation}
\tilde{\cal Z}(Z,Z^{*},B)
\equiv
\langle\, \mathcal{V}\mid B \, \rangle \, ,
\qquad
\tilde{\cal Z}(Z,Z^{*},B_{\pm})
\equiv
\langle\, \mathcal{V}\mid B_{\pm} \, \rangle \, ,
\end{equation}
\noindent
with the shifted coefficients
\begin{equation}
\label{eq:shiftedB}
B^{M}_{\pm}
\equiv
\lim_{\tau\rightarrow \mp\infty}\frac{r_{0}H^{M}(\tau)}{\sinh(r_{0}\tau)} =
B^{M}\mp r_{0} A^{M}\, .
\end{equation}

Imposing the same asymptotic conditions on the fields as in the extremal case
and the condition (\ref{eq:noNUTconstraintnonext}), we arrive again at
(\ref{eq:AMextremal}). Left to be determined from the equations of motion are
then only the constants $B^{M}$ and the non-extremality parameter $r_{0}$.

The mass is given again by eq.~(\ref{eq:massformulad4-1}) and the
expressions for the event horizon area ($+$) and the Cauchy horizon
area ($-$) are
\begin{eqnarray}
\label{eq:AWBnonext}
\frac{A_{\rm h\pm}}{4\pi}
=
\mathsf{W}(B_{\pm})\, .
\end{eqnarray}
\noindent
In the near-horizon limit, the equations of motion, upon use of the
above formulae for the area of the event horizon, lead to the
following relations
\begin{eqnarray}
\label{eq:AvsVhnonext}
\frac{A_{\rm h\pm}}{4\pi}
& = &
-V_{\rm bh}(B_{\pm})\pm 2r_{0}\mathcal{M}_{MN}[\mathcal{F}(B_{\pm})] A^{M} B^{N}_{\pm}
=
\mathsf{W}(B_{\pm})\, ,
\\
& & \nonumber \\
\label{eq:dV}
\partial_{P}V_{\rm bh}(B_{\pm})
& = &
\pm 2r_{0}\partial_{P}\mathcal{M}_{MN}[\mathcal{F}(B)] A^{M} B^{N}_{\pm}
=
- 2r^{2}_{0}\partial_{P}\mathcal{M}_{MN}[\mathcal{F}(B)] A^{M} A^{N}
\, ,
\end{eqnarray}
\noindent
which generalize eqs.~(\ref{eq:AvsVh4d}) and (\ref{eq:dVh=0}) to the
non-extremal case. In the last relation we have used the identity
\begin{equation}
H^{M}\partial_{P}\mathcal{M}_{MN}(\mathcal{F})=0\, .
\end{equation}

The right-hand side of eq.~(\ref{eq:dV}) vanishes if $A^{M} \propto
B^{M}$. This is a special case that we study in
section~\ref{sec:d4nonextremaldoublyextremal}. Another possibility is that
$\mathcal{F}_{\Lambda\Sigma}$ and hence also $\mathcal{M}_{MN}(\mathcal{F})$
are constant, as happens in quadratic models. In general, however,
$\partial_{P}V_{\rm bh}(B_{\pm}) \neq 0$ and we conclude that the values of the
scalars on the horizon of a non-extremal black hole do not necessarily
extremize the black-hole potential.


\subsubsection{First-order flow equations}
\label{First_order_NnExt}

The derivation carried out for extremal black holes in
section~\ref{First_order_ext} can be straightforwardly extended to the
non-extremal case. As in the 5-dimensional case studied in
ref.~\cite{Meessen:2012su}, one defines a new coordinate $\rho$ and a function
$f(\rho)$
\begin{equation}
\label{eq:almost-extremal}
\rho\equiv\frac{\sinh(r_{0}\tau)}{r_{0}\cosh(r_{0}\tau)}\,,
\qquad
f(\rho)\equiv\frac{1}{\sqrt{1-r_{0}^{2}\rho^{2}}}=\cosh(r_{0}\tau)\,,
\end{equation}
\noindent
so that the hyperbolic ansatz \eqref{eq:Ansatz} for $H^{M}$ can be
rewritten in the ``almost extremal form'':
\begin{equation}
H^{M}=f(\rho)(A^{M}+B^{M}\rho)\equiv f(\rho)\hat H^{M}\,.
\end{equation}
\noindent
Then, following the same steps that led to eqs.~\eqref{1st_order_Uext} and
\eqref{1st_Zhat}, one can obtain the first-order flow equations:
\begin{eqnarray}
 \label{1st_Uhat}
\frac{de^{-\hat{U}}}{d \rho}
& = &
\sqrt{2} |\tilde{\mathcal{\mathcal{Z}}}(Z,Z^{*},B)|\,,\\
\nonumber\\
\label{1st_Zhat}
\frac{d Z^{i}}{d \rho}
& = &
-2\sqrt{2}\,e^{\hat{U}}\mathcal{G}^{ij^{*}}
\partial_{j^{*}}|\tilde{\mathcal{Z}}(Z,Z^{*},B)|\,,
\end{eqnarray}
\noindent
where we have introduced the hatted warp factor $\hat{U}=U+\log{f}$.

Similarly to the extremal case, it is not difficult to show that this
first-order flow implies the second-order equations:
\begin{eqnarray}
\frac{d^{2} \hat{U}}{d\rho^{2}}
+
e^{2\hat{U}}V_{\mathrm{bh}}(Z,Z^{*},\sqrt{2}B)
& = &
0\,,\\
\nonumber\\
\frac{d^{2} Z^{i}}{d\rho^{2}}
+
\Gamma_{kl}{}^{i}\frac{d Z^{k}}{d\rho}\frac{d Z^{l}}{d\rho}
+
e^{2\hat{U}}\mathcal{G}^{ij^{*}}\partial_{j^{*}}V_{\mathrm{bh}}(Z,Z^{*},\sqrt{2}B)
& = &
0\, ,
\end{eqnarray}
\noindent
plus the constraint\footnote{Observe that the right-hand side of this equation
  is not $r_{0}^{2}$.}
\begin{equation}
\label{eq:jamiltonian}
\left(\frac{d \hat{U}}{d\rho}  \right)^{2}
+
\mathcal{G}_{ij^{*}}\frac{d Z^{i}}{d \rho}\frac{d Z^{*\, j^{*}}}{d \rho}
+
e^{2\hat{U}}V_{\mathrm{bh}}(Z,Z^{*},\sqrt{2}B)
=
0\, ,
\end{equation}
\noindent
but now with respect to the new variable $\rho$ and the new function $\hat{U}$.

In order to compare these equations with the actual second-order equations for
the warp factor and the scalars we have to rewrite them in terms of the
variable $\tau$ and rescale $\hat{U}$ to $U$. For the former, by using
$d/d\rho=f^{2}d/d\tau$ and eq.~\eqref{1st_Uhat}, one finds:
\begin{equation}
\ddot{U} -\frac{2\sqrt{2} \rho}{f} e^{U} |\mathcal{Z}(Z,Z^{*},\sqrt{2}B)|
+\frac{r_{0}^{2}}{f^{2}}
+\frac{e^{2U}}{f^{2}}
V_{\mathrm{bh}}(Z,Z^{*},\sqrt{2}B)\, ,
\end{equation}
\noindent
from which follows the relation between the true and the fake black hole
potential that must hold for the above second-order equations to imply the
equations of motion:
\begin{equation}
\label{U_VBH}
e^{2U}V_{\mathrm{bh}}(Z,Z^{*},\mathcal{Q})
=
\frac{e^{2U}}{f^{2}}
V_{\mathrm{bh}}(Z,Z^{*},\sqrt{2}B)
-\frac{2\sqrt{2} r_{0}^{2}\rho}{f} e^{U} |\mathcal{Z}(Z,Z^{*},\sqrt{2}B)|
+\frac{r_{0}^{2}}{f^{2}}\, .
\end{equation}
The same condition ensures that the constraint eq.~(\ref{eq:jamiltonian})
implies the standard Hamiltonian constraint. For the scalar equations we find
the condition
\begin{equation}
\partial_{i}
\left(
e^{2U}V_{\mathrm{bh}}(Z,Z^{*},\mathcal{Q})
-
\frac{e^{2U}}{f^{2}}
V_{\mathrm{bh}}(Z,Z^{*},\sqrt{2}B)
+\frac{4\sqrt{2} r_{0}^{2}\rho}{f} e^{U} |\mathcal{Z}(Z,Z^{*},\sqrt{2}B)|
\right)
=0\, .
\end{equation}
No other conditions need to be satisfied for the first-order equations to imply
all the second-order equations of motion. Taking the derivative with respect to
$\rho$ of eq.~(\ref{U_VBH}) we find that, if this relation is satisfied for any
$Z^{i}$ (or any $H^{M}$), then the last equation is also satisfied, as are all
the second-order equations.

Evaluating eq.~(\ref{U_VBH}) at spatial infinity ($\tau=0$, which corresponds
to $\rho=0$) we find the following relation between the charges, the fake
charges, the asymptotic values of the moduli and the non-extremality parameter:
\begin{equation}
\label{VBH_relation}
V_{\mathrm{bh}}(Z_{\infty},Z^{*}_{\infty},\mathcal{Q})
-V_{\mathrm{bh}}(Z_{\infty},Z^{*}_{\infty},\sqrt{2}B)
=
r_{0}^{2}\, .
\end{equation}


\subsubsection{Non-extremal generalization of doubly-extremal black holes}
\label{sec:d4nonextremaldoublyextremal}

For non-extremal black holes whose scalars are constant over the whole
spacetime, it is possible to solve the equations of motion of the H-FGK system
with the hyperbolic ansatz (\ref{eq:Ansatz}) in a model-independent way,
i.e.~for any theory of $N=2,d=4$ supergravity. Given the constancy of
the scalars we assume
\begin{equation}
\label{eq:scalarsiguales}
Z^{i}_{\infty} = Z^{i}_{\rm h}\, ,
\end{equation}
\noindent
which requires
\begin{equation}
B^{M} \propto A^{M}\, ,
\end{equation}
\noindent
where the constants $A^{M}$ are given by eq.~(\ref{eq:AMextremal}).

Using the proportionality of the $B^{M}$ and $A^{M}$ in the $\tau\rightarrow
0^{-}$ or $\tau\rightarrow \pm \infty$ limit of
eq.~\eqref{eq:equationsofmotion2nonext} we get
\begin{equation}
\label{eq:attractordoubly}
\partial_{K}V_{\mathrm{bh}}(Z_{\infty},Z^{*}_{\infty},\mathcal{Q}) = 0\, ,
\end{equation}
\noindent
which proves that the scalars must assume attractor values
$Z^{i}_{\infty}=Z^{i}_{\mathrm{att}}$ that are a stationary point of the black
hole potential, just as in the extremal case. We can thus use
eq.~(\ref{eq:AvsVh4d}), which gives the value of the black-hole potential at
the horizons in terms of the fake central charge there $\tilde{\cal Z}(B,B)$
(not $\tilde{\cal Z}(Z,Z^{*},B_{\pm})$):
\begin{equation}
\label{eq:VbhvsBdoubly}
-V_{\rm bh}(Z_{\infty},Z^{*}_{\infty},\mathcal{Q}) = |\tilde{\cal Z}(B,B)|^{2}\, .
\end{equation}

The proportionality constant between $B^{M}$ and $A^{M}$ is easily
determined to be $-\mathsf{W}^{1/2}(B)$ by using the normalization at
infinity $\mathsf{W}(A)=1$ and choosing the sign so as to make the
functions $H^{M}\neq 0$ for $\tau \in (-\infty,0)$. Then we can write
\begin{equation}
\label{eq:Bdoubly2}
H^{M}(\tau)
=
A^{M}
\left(
\cosh{(r_{0}\tau)}
-\mathsf{W}^{1/2}(B)\frac{\sinh{(r_{0}\tau)}}{r_{0}}
\right)\, .
\end{equation}
\noindent
The values of $B^{M}_{\pm}$ are
\begin{equation}
B^{M}_{\pm} = - \left(\mathsf{W}^{1/2}(B) \pm r_{0}\right)A^{M}\, ,
\end{equation}
\noindent
and
\begin{equation}
\mathsf{W}(B_{\pm})
=
\left(\mathsf{W}^{1/2}(B) \pm r_{0}\right)^{2}\, .
\end{equation}

A relation between the value of $\mathsf{W}^{1/2}(B)$ and physical parameters
and $r_{0}$ can be found by taking the $\tau\rightarrow 0^{-}$ limit of
eq.~\eqref{eq:hamiltonianconstraint2nonext}:
\begin{equation}
\label{eq:Wdoubly}
\mathsf{W}(B) = r_{0}^{2} -V_{\mathrm{bh}}(Z_{\infty},Z^{*}_{\infty},\mathcal{Q})\, .
\end{equation}
\noindent
Another relation comes from the definition of mass $M = \dot{U}(0)$,
which gives $M = -\tilde{H}_{M}(A)B^{M}$. Using the proportionality
between $A^{M}$ and $B^{M}$ we find that
\begin{equation}
M = \mathsf{W}^{1/2}(B)\, .
\end{equation}

The final expression for the functions $H^{M}(\tau)$ is, regardless of the
details of the model:
\begin{eqnarray}
H^{M}(\tau)
& = &
A^{M}
\left(
\cosh{(r_{0}\tau)}
-M\frac{\sinh{(r_{0}\tau)}}{r_{0}}
\right) ,
\\
& & \nonumber \\
\label{eq:entropy_doubly}
S_{\pm}
& = &
\pi\left(M \pm r_{0}\right)^{2}\, ,
\end{eqnarray}
\noindent
where the non-extremality parameter, upon use of
eq.~(\ref{eq:VbhvsBdoubly}), is given by
\begin{equation}
r_{0} = \sqrt{M^{2}-|\tilde{\cal Z}(B,B)|^{2}}\, .
\end{equation}


\section{One-modulus quantum-corrected geometries}
\label{sec:Example}


We shall now use the formalism developed in the last section to
explore the black-hole solutions of one-modulus quantum-corrected
models that typically appear as one-modulus Calabi--Yau
compactification of type II string theory. For one-modulus models of
this kind the perturbative prepotential $\mathcal{F}_{\rm pert}$ can
be brought to the form:
\begin{equation}\label{eq:prepCorrIIA}
\mathcal{F}^{\rm pert}_{\rm IIA}
=
-\frac{\kappa^{0}_{1,1,1}}{6} \frac{(\hat{\mathcal{X}}^{1})^{3}}{\hat{\mathcal{X}}^{0}}
-\frac{i}{2} c (\hat{\mathcal{X}}^{0})^{2}\, ,
\end{equation}
\noindent
where the \emph{correction} is encoded in the model-dependent positive
constant $c$, $\kappa^{0}_{1,1,1}$ is the triple intersection number
and the hat indicates that we are working in a possibly rotated (by a
symplectic matrix) frame of the homogeneous coordinates
$\{\mathcal{X}^0,\,\mathcal{X}^i\}$ of the moduli space. In what
follows we take the explicit example of the type~IIA superstring
compactified on the quintic Calabi--Yau manifold
($\kappa^{0}_{1,1,1}=5$), which we review in the appendix.

For the sake of simplicity and in order to be able to make a
comparison, in the following we first study the \textit{uncorrected}
model corresponding to the prepotential $\mathcal{F}^{0}_{\rm
  IIA}\equiv \mathcal{F}^{\rm pert}_{\rm IIA}(c=0)$ and only
afterwards the general case of eq.~\eqref{eq:prepCorrIIA}.


\subsection{Uncorrected case: the $t^{3}$ model}
\label{sec:uncorrectedt3}


In this section we consider the tree-level prepotential:
\begin{equation}
\mathcal{F}^{0}_{\rm pert}(\mathcal{X})
=
-\frac{5}{6}\frac{(\mathcal{X}^{1})^{3}}{\mathcal{X}^{0}}\, .
\end{equation}
\noindent
In terms of the coordinate $t=\mathcal{X}^{1}/\mathcal{X}^{0}$
the K\"ahler potential and metric are given by:
\begin{equation}
e^{-\mathcal{K}^{0}}
=
\tfrac{20}{3} (\Im \mathfrak{m}\, t)^{3}\, ,
\qquad
\mathcal{G}^{0}_{tt^{*}}
=
\tfrac{3}{4}\left(\Im\mathfrak{m}\, t\right)^{-2}\, ,
\end{equation}
\noindent
whereas the covariantly holomorphic symplectic section is
\begin{equation}
\label{eq:symplecticsectiont3}
\mathcal{V}^{0}(t,t^{*})
=
e^{\mathcal{K}^{0}/2}
\left(
\begin{array}{c}
1 \\
t \\
\frac{5}{6}t^{3} \\
-\frac{5}{2} t^{2} \\
\end{array}
\right)
\end{equation}
\noindent
and the central charge, its covariant derivative, the black-hole potential and
its partial derivative read:
\begin{eqnarray}
\label{eq:centralcharget3}
\mathcal{Z}
& \equiv &
e^{\mathcal{K}^{0}/2}
\hat{\mathcal{Z}}\, ,
\\
& & \nonumber \\
\label{eq:DtZ}
\mathcal{D}_{t}\mathcal{Z}
& = &
\frac{i}{2}
\frac{e^{\mathcal{K}^{0}/2}}{\Im\mathfrak{m}\, t}
\hat{\mathcal{W}}\, ,
\\
& & \nonumber \\
-V_{\rm bh}
& = &
e^{\mathcal{K}^{0}}
\left(|\hat{\mathcal{Z}}|^{2} +\tfrac{1}{3}|\hat{\mathcal{W}}|^{2}\right) ,
\\
& & \nonumber \\
\label{eq:dtV}
-\partial_{t}V_{\rm bh}
& = &
\tfrac{i}{20}
(\Im \mathfrak{m}\, t)^{-4}
\left( (\hat{\mathcal{W}}^{*})^{2}
+3 \hat{\mathcal{W}} \hat{\mathcal{Z}}^{*}\right) .
\end{eqnarray}
\noindent
In the above:
\begin{eqnarray}
\hat{\mathcal{Z}}
& = &
\tfrac{5}{6}p^{0} t^{3} -\tfrac{5}{2}p^{1}t^{2} -q_{1} t -q_{0}\, ,
\\
& & \nonumber \\
\label{eq:hatW}
\hat{\mathcal{W}}
& = &
\tfrac{5}{2}p^{0}t^{2}t^{*}
-\tfrac{5}{2}p^{1} t(t+2t^{*})
-q^{1}(2t+t^{*}) -3q^{0}\, .
\end{eqnarray}
\noindent
Notice all these objects are well defined only for $\Im\mathfrak{m}\, t >
0$. Furthermore, it must be taken into account that the theory given by the
tree-level prepotential is a good approximation to the full theory
only when $|t|\gg 1$.


\subsubsection{Extremal solutions}
\label{sec:extreme0}


Extremal solutions are associated with the critical points of the
black-hole potential. Following from eqs.~(\ref{eq:DtZ}) and
(\ref{eq:dtV}), there are two kinds of critical points:
\begin{enumerate}
\item Supersymmetric, when
\begin{equation}
\hat{\mathcal{W}}=0\, .
\end{equation}
\noindent
For generic (non-vanishing) values of the charges, there exist three
complex solutions for the critical values $t_{\mathrm{att}}$, but at most
two can be physical ($\Im\mathfrak{m}\, t >0$). Their expressions are
complicated and will be recovered below by taking the appropriate
limits in the solutions.

\item Non-supersymmetric \cite{Tripathy:2005qp,Saraikin:2007jc}, when
  $\hat{\mathcal{W}}\neq 0$ and
\begin{equation}
3\hat{\mathcal{Z}}\hat{\mathcal{W}}^{*}+\hat{\mathcal{W}}^{2}=0\, .
\end{equation}

\end{enumerate}

The extremal BPS solutions can be constructed by the procedure
explained in section~\ref{sec:d4extremal}. The Freudenthal duality equations
can be solved in a general way \cite{Shmakova:1996nz} and the metric
function and scalar field read:
\begin{equation}
\label{eq:generalsusysolutiont3}
  \begin{array}{rcl}
e^{-2U} & = &
\mathsf{W}(H)
=
\tfrac{2}{\sqrt{3}}\sqrt{\tfrac{8}{15} H^{0} (H_{1})^{3}
+(H^{1}H_{1})^{2}
-3(H^{0}H_{0})^{2}
-6H^{0}H_{0} H^{1} H_{1}
-10(H^{1})^{3} H_{0}
}\, ,
\\
& & \\
t
& = &
{\displaystyle
-\frac{3 H^{0}H_{0} + H^{1} H_{1}}{5 (H^{1})^{2} + 2 H^{0}H_{1}}
+i\frac{3e^{-2 U}}{2\left[5 (H^{1})^{2} + 2 H^{0} H_{1}\right]}\, .
}
\end{array}
\end{equation}
\noindent
The harmonic functions $(H^{M}) = (H^{0},H^{1},H_{0},H_{1})$ are given
by eq.~(\ref{eq:harmonicfunctions}) with $B^{M}=\mathcal{Q}^{M}$ and
the $A^{M}$ are given by eq.~(\ref{eq:AMextremal}) (with the upper
sign), where now the asymptotic values of the symplectic section
(\ref{eq:symplecticsectiont3}) and the central charge
(\ref{eq:centralcharget3}) have to be used. This guarantees the
absence of NUT charge (necessary for the consistency of the solution)
and the correct asymptotic behavior of the above fields:
$e^{-2U(0)}=1$, $t(0)=t_{\infty}$.

On the horizon, the values taken by these fields can be found by replacing the
harmonic functions $H^{M}$ by $-\mathcal{Q}^{M}/\sqrt{2}$, that is
\begin{equation}
  \begin{array}{rcl}
S_{\mathrm{e}}/\pi & = &
\tfrac{1}{2}\mathsf{W}(\mathcal{Q})
=
\frac{1}{\sqrt{3}}\sqrt{\tfrac{8}{15} p^{0} (q_{1})^{3}
+(p^{1}q_{1})^{2}
-3(p^{0}q_{0})^{2}
-6p^{0}q_{0}p^{1}q_{1}
-10(p^{1})^{3} q_{0}
}\, ,
\\
& & \\
t_{\mathrm{att}}
& = &
{\displaystyle
-\frac{3 p^{0}q_{0} + p^{1} q_{1}}{5 (p^{1})^{2} + 2 p^{0}q_{1}}
+i\frac{3\mathsf{W}(\mathcal{Q})}{2[5 (p^{1})^{2} + 2 p^{0} q_{1}]}\, .
}
\\
\end{array}
\end{equation}

The values of the fields on the horizon are well defined only if the
charges are such that the entropy and, hence,
$\mathsf{W}(\mathcal{Q})$ is real and non-vanishing and if
$\Im\mathfrak{m}\,t>0$. Furthermore, in order to be able to write the
above expressions we have assumed that $p^{0}>0$. Then, the conditions
that the charges must satisfy are
\begin{eqnarray}
\label{eq:ineqQ1}
p^{0}
& >& 0\, ,
\\
& & \nonumber \\
5(p^{1})^{2}+2p^{0}q_{1}
& > &
0\, ,
\\
& & \nonumber \\
\label{eq:ineqQ3}
\tfrac{8}{15} p^{0} (q_{1})^{3}
+(p^{1}q_{1})^{2}
-3(p^{0}q_{0})^{2}
-6p^{0}q_{0}p^{1}q_{1}
-10(p^{1})^{3} q_{0}
& > &
0\, .
\end{eqnarray}

The analysis of the possible values of the charges in the most general
case is complicated and unilluminating, so we will not attempt it
here. The inequalities (\ref{eq:ineqQ1})--(\ref{eq:ineqQ3}) must be
extended to the $H^{M}$ in order to guarantee the regularity of the
solution. The first-order flow equations imply that the metric
function grows monotonically from spatial infinity to the event
horizon, therefore it is enough to give it admissible values there to
ensure that it does not vanish for any value of $\tau \in (-\infty,
0)$. A similar argument applies to the scalar field.\footnote{With
  more scalar fields and non-diagonal metrics it would be more
  complicated to argue the same.}


Because the general supersymmetric solution turns out to be very
difficult to deform into the general non-extremal solution, we
consider a simpler three-charge case with $p^{0}=0$. The
supersymmetric solution (with $H^{0}_{\infty}=0$ as well) takes the
form:
\begin{equation}
\label{eq:susysolutiont3p0=0}
\begin{array}{rcl}
e^{-2U} & = &
\frac{2}{\sqrt{3}}|H^{1}|\sqrt{(H_{1})^{2} -10H^{1}H_{0}}\, ,
\\
& & \\
t
& = &
{\displaystyle
-\frac{H_{1}}{5H^{1}}
+i\frac{\sqrt{3}}{5}\frac{\sqrt{(H_{1})^{2} -10H^{1}H_{0}}}{|H^{1}|}\, ,
}
\end{array}
\end{equation}
%

For this simpler charge configuration it is also possible to directly
study the stationary points of the black hole potential to find a
non-supersymmetric critical point given by:
\begin{equation}
\label{eq:extremalnonsusysolutiont3p0=0}
t_{\mathrm{att}}
=
-\frac{q_{1}}{5p^{1}}
+i\frac{\sqrt{3}}{5}\frac{\sqrt{-[(q_{1})^{2} -10p^{1}q_{0}]}}{|p^{1}|}
\end{equation}
\noindent
and the corresponding entropy:
\begin{equation}
\label{eq:ENTOPYextremalnonsusysolutiont3p0=0}
S_{\mathrm{e}}/\pi =
\frac{1}{\sqrt{3}}|p^{1}|\sqrt{-[(q_{1})^{2} -10p^{1}q_{0}]}\, .
\end{equation}
\noindent
They differ from the supersymmetric case by the sign of the
discriminant 
\begin{equation}
\label{eq:discriminant}
\Lambda=  -p^{1}q_{0}+\frac{(q_{1})^{2}}{10}\, .
\end{equation}
\noindent
Rather than trying to construct the corresponding solutions directly,
we shall obtain them as a limit of the non-extremal solution that we
construct using the general procedure discussed in the previous
section.


\subsubsection{Non-extremal solution with $p^{0}=0$}
\label{sec:nonextreme0}


As we showed in section~\ref{sec:d4ansatz}, by using the ansatz
\begin{equation}
 H^{M}(\tau)=A^{M}\cosh{(r_{0}\tau)} + \frac{B^{M}}{r_{0}}\sinh{(r_{0}\tau)}\, .
\end{equation}
\noindent
valid for non-extremal black holes satisfying $H^{M}\dot H_M=0$, one
can reduce the differential equations of motion to the algebraic
equations
\eqref{eq:equationsofmotion2nonext}--\eqref{eq:noNUTconstraintnonext}
and solve them for the coefficients $B^{M}$. For a non-extremal black
hole in the $t^{3}$ model with charges $p^{1}$, $q_{0}$ and $q_{1}$
one finds:
\begin{eqnarray}
\label{B0}
B_{0}
& = &
 s^{1}\left(\sqrt{\frac{\Lambda^{2}}{2 (p^{1})^{2}}
+ \frac{5 r_{0}^{2} (\Im \mathfrak{m}\, t_{\infty})^{3}}{24}} -
\frac{q_{1}^{2}}{10(p^{1})^{2}} \sqrt{\frac{(p^{1})^{2}}{2}
+\frac{3r_{0}^{2}}{10 (\Im \mathfrak{m}\, t_{\infty})^{3}}}\right) ,
\\
& &\nonumber\\
\label{B1}
B^{1}
& = &
-s^{1}
\sqrt{\frac{3r_{0}^{2}}{10\Im \mathfrak{m}\, t_{\infty} }  +\frac12(p^{1})^{2}}\,,
\\& &\nonumber\\
\label{B11}B_{1}&=&
-s^{1}\frac{q_{1}}{p^{1}}
\sqrt{\frac{3r_{0}^{2}}{10\Im \mathfrak{m}\, t_{\infty}} +\frac12 (p^{1})^{2}}\,,
\end{eqnarray}
\noindent
where we have defined
\begin{equation}
s^{1}  \equiv 
\operatorname{sgn}(p^{1})\, ,
\end{equation}

The coefficients $A^{M}$ can be determined by using the general expression
\eqref{eq:AMextremal} and in our case turn out to be:
\begin{eqnarray}
A_{0}
& = &
s^{1}\frac{\sqrt{3}}{10\sqrt{10} \sqrt{\Im\mathfrak{m}\, t_{\infty}}}
\left[
\left(\frac{q_{1}}{p^{1}}\right)^{2}
-\frac{25}{3} \left(\Im\mathfrak{m}\, t_{\infty}\right)^{2}
\right]\, ,
\\
& & \nonumber \\
A^{1}
& = &
s^{1}\sqrt{\frac{3}{10 \Im \mathfrak{m}\,t_{\infty}}}\,,
\\
& & \nonumber \\
A_{1}
& = &
s^{1}\frac{q_{1}}{p^{1}}\sqrt{\frac{3}{10 \Im \mathfrak{m}\,t_{\infty}}}\, .
\end{eqnarray}

From the relation $M= \dot{U}(0)$ the mass is found to be
\begin{equation}
\begin{split}
M =
\frac{1}{4}
\Biggl(&\sqrt{\frac{-60 p^{1} q_{0} (q_{1})^{2} + 3 (q_{1})^{4}
+ 25 (p^{1})^{2} [12 (q_{0})^{2}
+ 5 r_{0}^{2} (\Im \mathfrak{m}\, t_{\infty})^{3}]}{125 (p^{1})^{2}
(\Im\mathfrak{m}\,t_{\infty})^{3}}}\\
&+\sqrt{9 r_{0}^{2}
+ 15 (p^{1})^{2} (\Im \mathfrak{m}\, t_{\infty})^{3}}\Biggr) .
\end{split}
\end{equation}
\noindent
One can invert this expression to obtain $r_{0}$ in terms of the
physical parameters $M$, $\Im \mathfrak{m}\,t_{\infty}$, $p^{1}$,
$q_{0}$:
\begin{equation}
\begin{split}
r_{0}^{2}
={}& \frac{1}{1000 (p^{1})^{4} \Im \mathfrak{m}\,t_{\infty}^{6}}
\Bigl(-60 (p^{1})^{3} q_{0} (q_{1})^{2} \Im \mathfrak{m}\,t_{\infty}^{3} + 3
  (p^{1})^{2} (q_{1})^{4} \Im \mathfrak{m}\,t_{\infty}^{3} \\
& - 1875 (p^{1})^{6} \Im \mathfrak{m}\,t_{\infty}^{7} + 100 (p^{1})^{4}
\left[3 (q_{0})^{2} \Im \mathfrak{m}\,t_{\infty}^{3} + 25 M^{2} \Im
\mathfrak{m}\,t_{\infty}^{6}\right] \\
& + 10 \sqrt{30M^{2} (p^{1})^{6} \Im \mathfrak{m}\,t_{\infty}^9 }\\
&\sqrt{9(q_{1})^{2}\left[(q_{1})^{2} - 20 p^{1} q_{0}\right]
+ 25 (p^{1})^{2} \left[36 (q_{0})^{2} - 25(p^{1})^{2}
\Im \mathfrak{m}\,t_{\infty}^{4} +
30 M^{2} \Im \mathfrak{m}\,t_{\infty}^{3}\right]}\Bigr).
\end{split}
\end{equation}

\begin{table}
\centering
\begin{tabular}{c|c|c}
$s^{1}$ & $s_{0}$ & $s_\Lambda$\\
\hline
$+$ & $-$ & $+$ \\
$-$ & $+$ & $+$ \\
$+$ & $+$ & $+$ \\
$-$ & $-$ & $+$ \\
$+$ & $+$ & $-$ \\
$-$ & $-$ & $-$ \\
\end{tabular}
\caption{The extremal limits depend on $s^{1}$ and $s_\Lambda$.
  Here $s_{0}=\operatorname{sgn}(q_{0})$, $s^{1}=\operatorname{sgn}(p^{1})$ and 
  ($s_{\Lambda}=\operatorname{sgn}(\Lambda)$ where the discriminant $\Lambda$
  has been defined in eq.~(\ref{eq:discriminant}).
  There are 6 possible cases :
  the first 4 possibilities ($s_\Lambda=+1$) would produce a supersymmetric
  extremal black hole while the others ($s_\Lambda=-1$) a
  non-supersymmetric one.}
\label{tab_signs}
\end{table}

This result allows one to obtain the expression for the mass in the
extremal limit $r_{0}\to 0$, namely:
\begin{equation}
\label{mass3c_ext}
M=
\sqrt{\frac35}\,\frac{25 (p^{1})^{2} \Im
  \mathfrak{m}\,t_{\infty}^{2}\!+ 10 \lvert\Lambda\rvert}{20 \lvert p^{1}\rvert \Im \mathfrak{m}\,t_{\infty}^{3/2}}\,.
\end{equation}
\noindent
It is easy to check that $M>0$. As mentioned at the end of the
previous section, when $s_\Lambda=\operatorname{sgn}(\Lambda)$ is
positive, the solution is supersymmetric (see table \ref{tab_signs}),
in which case the anharmonic function $H_{0}=A_{0}\cosh
(r_{0}\tau)+\tfrac{B_{0}}{r_{0}}\sinh(r_{0}\tau)$ becomes for
$r_{0}\to 0$:
\begin{equation}
H_{0}=s^{1}\frac{\sqrt{3}}{10\sqrt{10} \sqrt{\Im\mathfrak{m}\, t_{\infty}}}
\left[
\left(\frac{q_{1}}{p^{1}}\right)^{2}
-\frac{25}{3} \left(\Im\mathfrak{m}\, t_{\infty}\right)^{2}
\right]-\frac{1}{\sqrt{2}}q_{0}\tau\,,
\end{equation}
\noindent
whereas in the non-supersymmetric case:
\begin{equation}
H_{0}=s^{1}\frac{\sqrt{3}}{10\sqrt{10} \sqrt{\Im\mathfrak{m}\, t_{\infty}}}
\left[
\left(\frac{q_{1}}{p^{1}}\right)^{2}
-\frac{25}{3} \left(\Im\mathfrak{m}\, t_{\infty}\right)^{2}
\right]+\frac{1}{\sqrt{2}}\left( q_{0}-2\frac{q_{1}^{2}}{10p^{1}}\right)\tau\,.
\end{equation}
\noindent
The extremal limit for $H^{1}=\frac{p^{1}}{q_{1}}H_{1}$ is in turn:
\begin{equation}
H^{1}
=
s^{1}\sqrt{\frac{3}{10\Im \mathfrak{m}\, t_{\infty}}}-\frac{1}{\sqrt{2}}p^{1}\tau\,.
\end{equation}
\noindent
Accordingly, for the warp factor after some simplification one obtains
\begin{equation}
e^{-2U}=
\tfrac{2}{\sqrt{3}}\sqrt{\pm\left[-10(H^{1})^{3} H_{0} +(H^{1}H_{1})^{2}\right]
}\, ,
\end{equation}
\noindent
where the plus holds for supersymmetric solutions and the minus for
non-supersymmetric.

The entropies associated with the outer ($\tau\to-\infty$) and inner
($\tau\to+\infty$) horizon can be computed to be respectively:
\begin{eqnarray}
 \frac{S_+}{\pi}&=&\frac{1}{15^{3/4}}\left[\frac{\big(\sqrt{3} r_{0} + \sqrt{
    3 r_{0}^{2} + 5 (p^{1})^{2} \Im \mathfrak{m}\,t_{\infty}}\big)^{3}}{\Im \mathfrak{m}\,t_{\infty}^{2}} \right.
\\\nonumber
& &\left.\left(5 \sqrt{5} r_{0}+ \sqrt{
    \frac{300(q_{0})^{2}}{\Im \mathfrak{m}\,t_{\infty}^{3}} - \frac{60 q_{0} (q_{1})^{2}}{p^{1}\Im \mathfrak{m}\,t_{\infty}^{3}} + \frac{3 (q_{1})^{4}}{\Im \mathfrak{m}\,t_{\infty}^{3}(p^{1})^{2}} + 125 r_{0}^{2}}\right)\right]^{1/2}\,,
\\& &\nonumber\\
\frac{S_-}{\pi}&=&\frac{1}{15^{3/4}}\left[\frac{\big(-\sqrt{3} r_{0} + \sqrt{
    3 r_{0}^{2} + 5 (p^{1})^{2} \Im \mathfrak{m}\,t_{\infty}}\big)^{3}}{\Im \mathfrak{m}\,t_{\infty}^{2}} \right.
\\\nonumber
& &\left.\left(5 \sqrt{5} r_{0}- \sqrt{
    \frac{300(q_{0})^{2}}{\Im \mathfrak{m}\,t_{\infty}^{3}} - \frac{60 q_{0} (q_{1})^{2}}{p^{1}\Im \mathfrak{m}\,t_{\infty}^{3}} + \frac{3 (q_{1})^{4}}{\Im \mathfrak{m}\,t_{\infty}^{3}(p^{1})^{2}} + 125 r_{0}^{2}}\right)\right]^{1/2}\,.
\end{eqnarray}
\noindent
By taking the limit $r_{0}\to0$ the extremal black hole entropy is
recovered from both $S_+$ and $S_-$ and their product satisfies the
geometric mean property $S_+S_- = \tfrac{\pi^{2}}{3}(p^{1})^{2}\left[-10
p^{1} q_{0} + (q_{1})^{2}\right] = S_{\mathrm{e}}^{2}$.


\subsection{Quantum-corrected case}
\label{sec:quantumcorrected}


For the quantum-corrected model of type IIA superstring on the
quintic, whose prepotential can be brought to the form
\eqref{eq:prepCorrIIA} by a symplectic rotation of the coordinate
frame (see the appendix), the covariantly holomorphic period vector
reads:
\begin{equation}
\label{eq:VPert}
\mathcal{V}_{\rm pert}
=
e^{\mathcal{K}_{\rm pert}/2}\left(
\begin{array}{c}
1 \\
t \\
\tfrac{5}{6}t^{3}
-i c
\\
-\tfrac{5}{2}t^2
\\
\end{array}
\right) ,
\end{equation}
\noindent
where (in the compactification we are considering)
$c=\frac{25}{\pi^{3}}\zeta(3)\approx 0.969204$. Because the general
case is very complicated, we deal only with two-charge and
three-charge black holes.


\subsubsection{Supersymmetric solution with $\hat{\mathcal{Q}}=
  (\hat{p}^{0},0,0,\hat{q}_{1})^{\rm T}$, $\mathcal{Q}=
  (p^{0},0,0,q_{1})^{\rm T}$
}\label{sec:p0q1}

The relations between the two pairs of charges in the rotated frame and
in the original one are:
\begin{equation}
 \hat{p}^{0}=p^{0}\,,\qquad \hat{q}_{1}=q_{1}-\frac{25}{12} p^{0}\,.
\end{equation}
\noindent
By solving the equation for the extremal supersymmetric case one
finds:\footnote{As the $H^M$ in the original frame do not appear (and
  $\hat{H}^{M}$ has been already used with a different meaning in
    eq.~\eqref{eq:almost-extremal}), we suppress the hats on the
    rotated $H^M$.}
\begin{eqnarray}
\label{2c_susy_cor}
t&=&s_{i} i\sqrt{\frac25\frac{H_{1}}{H^{0}}}\,,\\
\nonumber\\
e^{-2U_{\rm e}}&=&s_{i}\frac43\sqrt{\frac25 H^{0} (H_{1})^{3}} + c (H^{0})^{2}\,,
\end{eqnarray}
\noindent
with $H^{M}=A^{M}-\tfrac{1}{\sqrt{2}}\hat{\mathcal{Q}}^{M}\tau$ and
$s_{i}=+1$ when
\begin{equation}\label{eq:2ch1}
\sqrt{\frac25\,\frac{\hat{q}_{1}}{\hat{p}^{0} }}\,\in\,\left(\Big(\frac{3c}{5}\Big)^{1/3},\infty\right),
\end{equation}
\noindent
while $s_{i}=-1$ for
\begin{equation}\label{eq:2ch2}
\sqrt{\frac25\,\frac{\hat{q}_{1}}{\hat{p}^{0} }}\,\in\,\left(0,\Big(\frac{3c}{10}\Big)^{1/3}\right),
\end{equation}
\noindent
so that $\Im\mathfrak{m}\,t$ lies in the allowed domain
\eqref{eq:tconstraint2} (for other values of the charges the
supersymmetric solution simply does not exist). By using
\eqref{eq:AMextremal} one can determine the constant part of the
harmonic functions:
\begin{equation}\label{A_nnExt_cor}
A^{0}=s_\mathcal{Q}\, \sqrt{\frac{3}{10\,s_{i}\,\Im
     \mathfrak{m}\,t_{\infty}^{3}+3c}}\,,\qquad
A_{1}=s_\mathcal{Q}\,\frac52\,\Im \mathfrak{m}\,t_{\infty}^{2}\sqrt{\frac{3}{10\,s_{i}\,\Im \mathfrak{m}\,t_{\infty}^{3}+3c}}\,.
\end{equation}

Notice that two disconnected branches of supersymmetric solutions
appear and only one of them, the case \eqref{eq:2ch1}, survives when
$c=0$. For both supersymmetric possibilities
$\operatorname{sgn}(\hat{p}^{0})=s_\mathcal{Q}=\operatorname{sgn}(\hat{q}_{1})$
and depending on the charges, the scalar at infinity is bound to a
certain set of possible values. If the charges, for example, satisfy
\eqref{eq:2ch1}
also $\Im \mathfrak{m}\,t_{\infty}$ must belong to this interval and
all the flow of the scalar in the moduli space takes place inside this
confined region. By looking at the explicit form of the solutions it
is possible to convince oneself that the two distinct branches of
solutions cannot be connected smoothly by changing the value of the
charges.

The entropy and the mass, once computed, can be written in the form:
\begin{eqnarray}
\frac{S_\mathrm{e}}{\pi}&=&\frac{\tfrac{45}{4}c^{2}(\hat{p}^{0})^{3}+8(\hat{q}_{1})^{3}+s_{i}\,6\sqrt{10}\,c\,\sqrt{(\hat{p}^{0} \hat{q}_{1})^{3}}}{\tfrac{45}{2}c\,\hat{p}^{0}+ s_{i}\,6\sqrt{10}\,\hat{q}_{1} \sqrt{\frac{\hat{q}_{1}}{\hat{p}^{0}}}}\,,\\
\nonumber\\
M_\mathrm{e}&=&\frac{\left|6c\,\hat{p}^{0}+6\,\hat{q}_{1} \Im \mathfrak{m}\,t_{\infty}+5\, \hat{p}^{0} \Im \mathfrak{m}\,t_{\infty}^{3}\right|}{4\sqrt{\tfrac{9}{2}c+15 \Im \mathfrak{m}\,t_{\infty}^{3}}}\,.
\end{eqnarray}
\noindent
The positivity of both the entropy and the mass is guaranteed by the fact that
the charges are confined to the intervals \eqref{eq:2ch1}, \eqref{eq:2ch2}.

The study of this two-charge configuration in the rotated symplectic
frame allows the analysis of the single charge configurations
$\mathcal{Q}=(p^{0},0,0,0)^{\rm T}$ and
$\mathcal{Q}=(0,0,0,q_{1})^{\rm T}$ in the original frame. For the
former one should substitute in the formulae above
$\hat{q}_{1}=-\frac{25}{12}\hat{p}^{0}$ but already here an
inconsistency occurs due to the requirement
$\operatorname{sgn}(\hat{p}^{0})=\operatorname{sgn}(\hat{q}_{1})$ that
would not be respected. Also for the other single-charge
configuration, by setting $\hat{p}^{0}=p^{0}=0$, it is easy to realize
that the expressions become ill-defined. This suggests that no
physical BPS solutions exist for the single-charge case at hand.

Before passing to non-extremal black holes, it is worth mentioning
that the Freudenthal duality equations also admit a solution that cannot be
accepted, namely:
\begin{eqnarray}
t&=&\frac{3c}{2H_{1}}\,\sqrt{\frac{8}{45c^{2}}\frac{H^{3}_{1}}{H^{0}}-(H^{0})^{2}}
+ ic\frac{3H^{0}}{2H_{1}}\,,\\
\nonumber\\
e^{-2U_{\rm e}}&=&2(H^{0})^{2}c+\frac{2}{45c}\frac{(H_{1})^{3}}{H^{0}}\,.
\end{eqnarray}
\noindent
These expressions would be well defined only for charges that violate
the constraint \eqref{eq:tconstraint2}, which leads to invalid
K\"ahler metric and K\"ahler potential.


\subsubsection{Supersymmetric solution with $\hat{\mathcal{Q}}=
  (0,\hat{p}^{1},\hat{q}_{0},\hat{q}_{1})^{\rm T}$, $\mathcal{Q}=
  (0,p^{1},q_{0},q_{1})^{\rm T}$}


This configuration corresponds to a three-charge black hole also in the
original frame, according to the relations:
\begin{equation}\label{eq:3chRel}
\hat{p}^{1}=p^{1}\,,\qquad\hat{q}_{0}=q_{0}-\frac{25}{12} p^{1}\,,
\qquad\hat{q}_{1}=q_{1}+\frac{11}{2} p^{1}\,.
\end{equation}
\noindent
We solve the Freudenthal duality equations with the harmonic function
$H^{0}$ set to zero. This yields:
\begin{eqnarray}
\hat{\mathcal{X}}^{0}
& = &
\frac{\rho^{2}+\rho \alpha^{1/3}+\alpha^{2/3}}{30c\,H^{1}\,\alpha^{1/3}}\,,\\
\nonumber\\
\hat{\mathcal{X}}^{1}
& = &
i H^{1}-\frac{H_{1}}{5H^{1}}\hat{\mathcal{X}}^{0}\,,\\
\nonumber\\
\label{UExtrCorr}
U
& = &
-\frac1
2\log\left(\frac{\alpha^{1/3}\,\big(\beta+\gamma\alpha^{1/3}\big)+\delta}{100\,c\,(H^{1})^{2}\,\alpha^{2/3}\big(\alpha^{2/3}+\rho\,\alpha^{1/3}+\rho^{2}\big)}\right),
\end{eqnarray}
\noindent
where
\begin{equation}
\begin{split}
\rho&=-10H^{1} H_{0}+(H_{1})^{2}\,,\\
\\
\alpha&=\rho^{3}-11250c^{2}\,(H^{1})^{6}+150\,\sqrt{(H^{1})^{6}c^{2}[5625c^{2}(H^{1})^{6}-\rho^{3}]}\,,\\
\\
\beta&= \rho^{2} \left(\rho^{3}-7500 c^{2}\,(H^{1})^{6}+
   100\,\sqrt{(H^{1})^{6}c^{2}[5625c^{2}(H^{1})^{6}-\rho^{3}]}\right)\,,  \\
\\
\gamma&=\rho\left(\rho^{3}-3750 c^{2}\, (H^{1})^{6} +
   50\,\sqrt{(H^{1})^{6}c^{2}[5625c^{2}(H^{1})^{6}-\rho^{3}]}\right)\,, \\
\\
\delta&=\big(\rho^{3}+7500c^{2}\, (H^{1})^{6} \big) \alpha\,.\\
\end{split}
\end{equation}

The expression for the physical scalar then becomes
\begin{equation}\label{eq:scalarExtrCorr}
t=\frac{\hat{\mathcal{X}}^{1}}{\hat{\mathcal{X}}^{0}}=-\frac{\hat{q}_{1}}{5\hat{p}^{1}}+ i\,\frac{30c\,(H^{1})^{2}\alpha^{1/3}}{\alpha^{2/3}+\rho\,\alpha^{1/3}+\rho^{2}}\,.
\end{equation}
\noindent
The constant parts of the harmonic functions turn out to be:
\begin{align}
\label{eq:constExCor}
A^{1}&=s^{1}\,\frac{\sqrt{3}\Im \mathfrak{m}\,t_{\infty}}{\sqrt{3c+10\Im \mathfrak{m}\,t_{\infty}^{3}}}\,,\qquad A_{1}=s_{1}\,\frac{\sqrt{3}\hat{q}_{1}\,\Im \mathfrak{m}\,t_{\infty}}{\hat{p}^{1}\sqrt{3c+10\Im \mathfrak{m}\,t_{\infty}^{3}}}\,,\\
\nonumber\\
A_{0}&=s_{0}\,\frac{3(\hat{q}_{1})^{2}\Im \mathfrak{m}\,t_{\infty}-25(\hat{p}^{1})^{2}\Im \mathfrak{m}\,t_{\infty}^{3}-30c\,(\hat{p}^{1})^{2}}{10(\hat{p}^{1})^{2}\sqrt{9c+30\Im \mathfrak{m}\,t_{\infty}^{3}}}\,,
\end{align}
\noindent
where $s^M=\operatorname{sgn}(\hat{\mathcal{Q}}^{M})$.

The solution just displayed is a purely ``quantum black hole'': it
diverges when $c$ is put to zero and it is well defined only for a
restricted set of values of the parameters $\{\hat{p}^{1},\,\hat
q_{0},\,\hat{q}_{1},\, \Im \mathfrak{m}\,t_{\infty}\}$.  By looking at
the expressions of the scalar and the warp factor we realize that the
problematic part is the square root
\begin{equation}\label{eq:rootCond}
\sqrt{(H^{1})^{6}c^{2}[5625c^{2}(H^{1})^{6}-\rho^{3}]}
\end{equation}
\noindent
that, in order to be real, needs the radicand to be bigger than or equal to
zero. This condition must be considered besides the requirement that the
imaginary part of the scalar should belong to the intervals
\eqref{eq:tconstraint2} and the positivity of the warp factor. Then the
allowed values of the charges can be determined by studying the behavior of
solutions on the horizon whereas the allowed values for $\Im
\mathfrak{m}\,t_{\infty}$ are given by the limit at infinity ($\tau \to
0^-$). In the end one obtains the following restrictions:
\begin{gather}
 \label{eq:yinInterval}\Im \mathfrak{m}\,t_{\infty}\in\left(-\Big(\frac{3c}{10}\Big)^{1/3},0\right)\approx\Big(-0.662489\,,\,0\Big),\\
\nonumber\\
 \label{eq:q0>0}
\hat{q}_{0}>\frac{(\tfrac{75}{2}\,c)^{2/3}\,(\hat{p}^{1})^{2}+(\hat{q}_{1})^{2}}{10 \hat{p}^{1}}\quad \text{if}\quad \hat{p}^{1}>0\,,\\
\nonumber\\
 \label{eq:q0<0}
\hat{q}_{0}<\frac{(\tfrac{75}{2}\,c)^{2/3}\,(\hat{p}^{1})^{2}+(\hat{q}_{1})^{2}}{10 \hat{p}^{1}}\quad \text{if}\quad \hat{p}^{1}<0\,.
\end{gather}

It is not difficult to see that the conditions \eqref{eq:q0>0},
\eqref{eq:q0<0} would be violated by the charge configuration
$\hat{\mathcal{Q}}=(0,\hat p^{1},0,\hat{q}_{1})^{\rm T}$, which would
produce a black hole with singular metric (differently from the
uncorrected $t^{3}$ model). Similarly one can exclude the existence of
black holes with the charge vector $\hat{\mathcal{Q}}=(0,\hat
p^{1},-\frac{25}{12}\hat{p}^{1},\frac{11}{2}\hat{p}^{1})^{\rm T}$,
corresponding in the original frame to $\mathcal{Q}=(0,p^{1},0,0)$:
when $\hat{p}^{1}=p^{1}=0$ the expression for the scalar would
diverge. This last observation, together with the discussion in the
previous subsection, indicates that this model does not admit regular
supersymmetric single-charge black holes.

On the other hand, solutions with $H_{1}=0$ (corresponding to the
charge configuration
$\hat{\mathcal{Q}}=(0,\hat{p}_{1},\hat{q}_{0},0)^{\rm T}$,
$\mathcal{Q}=(0,p^{1},q_{0},-\frac{11}{2} p^{1})^{\rm T}$) or with
$q_{1}=0$ (two-charge in the unrotated frame), are physical. In the
former case the scalar becomes purely imaginary
\begin{equation}
\begin{split}
t=& -3 i\, \frac{(H^{1})^{2} c\, \lambda^{1/3}}{(H^{1})^{2} (H_{0})^{2} +H^{1} H_{0} \lambda^{1/3}+
    \lambda^{2/3}}\,,\\
&\\
\lambda=&\frac{45}{4} (H^{1})^{6} c^{2} + (H^{1})^{3} H_{0}^{3} -
      3  \sqrt{\frac{5}{4}(H^{1})^9 c^{2} \left(\frac{45}{4} (H^{1})^{3} c^{2} + 2 (H_{0})^{3}\right)}
\end{split}
\end{equation}
\noindent
and in line with eqs.~\eqref{eq:q0>0}, \eqref{eq:q0<0} the charges
must satisfy
$\operatorname{sgn}\hat{p}^{1}=\operatorname{sgn}\hat{q}_{0}$ and
$\left|\hat
  q_{0}\right|>(\frac{(75/2\,c)^{2/3}}{10})\left|\hat{p}^{1}\right|$. When
instead $q_{1}=0$, the real part of the scalar takes a fixed value
independent of parameters, namely $\Re\mathfrak{e}\,t=-\frac{11}{10}$,
and the restrictions on the allowed charges become
$\operatorname{sgn}\hat{q}_{0} = \operatorname{sgn}\hat{p}^{1}$ and
$\left|\hat
  q_{0}\right|>\frac{4(75/2\,c)^{2/3}+121}{40}\left|\hat{p}^{1}\right|$.

The entropy and the mass for the black holes in this section can be
calculated as usual, but due to the complexity of the expressions, we
do not display them.


\subsubsection{Non-extremal solutions}\label{sec:Non-extCorr}


The expressions for the scalar and warp factor are in general very
involved and this turns out to make the pursuit of non-supersymmetric
black holes cumbersome. The difficulty resides in the fact that
the equations for the coefficients turn out to be polynomials of a
very high degree, which cannot be solved analytically.

The only non-extremal black holes that can be quite straightforwardly
studied are those with the scalar assuming a constant value that
extremizes the black hole potential. From the general treatment in
\ref{sec:d4nonextremaldoublyextremal} we know that for such
non-extremal solutions 
\begin{equation}
B^{M}=-\,A^{M} M=-\,A^{M}\sqrt{\lvert
  \mathcal{Z}(Z_{\infty},Z^{*}_{\infty},\mathcal{Q})\rvert^{2}+r_{0}^{2}}\, ,
\end{equation}
\noindent
and the only quantity to calculate is the absolute value of the
central charge in the stationary points of the black hole
potential. In the current case it reads:
\begin{equation}\label{mdZ_nnExt_cor}
\lvert \mathcal{Z}(Z_{\infty},Z^{*}_{\infty},\mathcal{Q})\rvert=\frac{\lvert 6 \hat{q}_{0} + 6 \hat{q}_{1}t_\mathrm{att} + 15 \hat{p}^{1} t_\mathrm{att}^{2} - 5 \hat{p}^{0} t_\mathrm{att}^{3}+6 i c \hat{p}^{0}\rvert}{\sqrt{6(12c + 5 i \Im \mathfrak{m}\,t_\mathrm{att}^{3})}}\,,
\end{equation}
\noindent
where $t_\mathrm{att}$ is the constant value of the scalar all along the
flow.

So far no analytic expressions for non-supersymmetric stationary
points of $V_{\mathrm{bh}}(Z,Z^{*},\mathcal{Q})$ have been obtained for a
general charge configuration.\footnote{An accurate numerical study has
  been carried out in \cite{Bellucci:2007eh}.} We study the
non-extremal version of (some of) the supersymmetric black holes of
the previous subsections and present an example of a constant-scalar
non-extremal black hole built from a non-supersymmetric critical
point of a system with a particular charge vector.


\paragraph{Configuration $\hat{\mathcal{Q}}=
  (\hat{p}^{0},0,0,\hat{q}_{1})^{\rm T}$:}
When $\hat{q}_{1}\hat{p}^{0}>0$, we read off from
eq.~\eqref{2c_susy_cor} that
\begin{equation}
 t_{\mathrm{att}}=i\,s_{i}\sqrt{\frac{2\hat{q}_{1}}{5\hat{p}^{0}}}
\end{equation}
\noindent
and by plugging it in \eqref{A_nnExt_cor} and \eqref{mdZ_nnExt_cor} we find:
\begin{equation}
 B^{0}=-s_\mathcal{Q}\sqrt{\frac{15\,M^{2}}{15\,c+s_{i}\, 4\sqrt{10\left(\frac{\hat{q}_{1}}{\hat{p}^{0}}\right)^{3}}}}\,,
\qquad
B_{1}=\frac{\hat{q}_{1}}{\hat{p}^{0}}B^{0}\,.
\end{equation}
\noindent
where the mass $M$ is equal to:
\begin{equation}
 M=\sqrt{\frac{c}{2}\,(\hat{p}^{0})^{2}+s_{i}\sqrt{\frac{8}{45}\,\hat {p}^{0}(\hat{q}_{1})^{3}}+r_{0}^{2}} \,.
\end{equation}
\noindent With this last expression the outer and the inner entropy
follow from eq.~\eqref{eq:entropy_doubly}. It is worth noticing that all
these formulae reduce to the extremal counterparts in the limit
$r_{0}\to 0$ and for the entropy it holds $S_+ S_-=S_\mathrm{e}^{2}$.


\paragraph{Configuration
  $\hat{\mathcal{Q}}=(0,\hat{p}^{1},2\hat{p}^{1},0)^{\rm T}$:} For the
sake of simplicity we take $\hat{q}_{0}=2\hat{p}^{1}$. The black hole
potential has a charge-independent critical point (corresponding to a
supersymmetric attractor) at:
\begin{equation}
\begin{split}
t_{\mathrm{att}}
&
=-6\,i\,c\, \frac{\left(64 + 90\, c^{2} - 6c\sqrt{5\, (64 + 45 c^{2})}\,\right)^{1/3}}{
 12+\left(2+ \left(64 + 90 c^{2} - 6c\sqrt{5\, (64 + 45 c^{2})}\right)^{1/3}\right)^{2}}\equiv-6\,i\,c\,\xi\\
\nonumber \\
&\approx-0.447310\,i
\end{split}
\end{equation}
\noindent
and the coefficients of the hyperbolic functions are:
\begin{equation}
B^{1}=s^{1}\,\frac{6\,c\,M\,\xi}{\sqrt{c-720\,c^{3}\,\xi^{3}}}\,,\qquad
B_{0}=\frac{1-180\,c^{2}\,\xi^{3}}{6\,\xi}\,B^{1}\,.
\end{equation}
For the mass one finds:
\begin{equation}
M=\sqrt{(p^{1})^{2} \,\frac{2\,(1 - 45\, c^{2}\, \xi^{2})^{2}}{c\,(1 - 720 \,c^{2}\, \xi^{3})}+r_{0}^{2}}\,.
\end{equation}
\noindent
From these expressions it is easy to see by setting $c=0$ that this
black hole does not reduce to a regular solution of the $t^{3}$ model.


\paragraph{Configuration $\hat{\mathcal{Q}}=(\hat
  p^{0},0,0,-\tfrac{35}{2}(\tfrac32c)^{2/3}\hat{p}^{0})^{\rm T}$:}
Also in this case the stationary point of the black-hole potential
does not depend on the value of $\hat{p}^{0}$ (although this time it
corresponds to a non-supersymmetric attractor):
\begin{equation}
t_{\mathrm{att}}=i\left(\tfrac{3}{2}c\right)^{1/3}\approx 1.13284\,i\,.
\end{equation}
\noindent
The non-extremal solution with a constant scalar is then completely
characterized by
\begin{equation}
B^{0}= -\,s^0\,\frac{M}{\sqrt{6\,c}}\,,
\qquad
B_{1}=-\,s^0\,\frac{5}{4}\left(\frac32\,c\right)^{1/6}M\,,
\end{equation}
\noindent
with
\begin{equation}
 M=\sqrt{48\,c\,(\hat{p}^{0})^{2}+r_{0}^{2}}\,.
\end{equation}
\noindent
The limit $r_{0}\to 0$ gives a doubly-extremal non-supersymmetric
black hole. Setting $c=0$ again does not lead to a regular solution.


\paragraph{Configurations $\hat{\mathcal{Q}} =
  (0,\hat{p}^{1},0,0)^\mathrm{T}$ and $\hat{\mathcal{Q}} =
  (0,0,0,\hat{q}_{1})^\mathrm{T}$:} Of these two configurations that
are both single-charge in the rotated frame, the second is one-charge
also in the original frame, $\mathcal{Q} =
(0,0,0,\hat{q}_{1})^\mathrm{T} = (0,0,0,q_{1})^\mathrm{T}$. The
admissible critical points of the black hole potential
$-V_{\mathrm{bh}}$ give in each case one non-supersymmetric attractor,
\begin{equation}
\label{eq:tcrph1}
t_{\mathrm{att}} =
i\sqrt[3]{\left(6+\sqrt[3]{206-6\sqrt{87}}+\frac{17\sqrt[3]{4}}{\sqrt[3]{103-3\sqrt{87}}}\right)\frac{3c}{10}}
\approx 1.37065\,i
\end{equation}
\noindent
or
\begin{equation}
t_{\mathrm{att}} = -i\sqrt[3]{(3\sqrt{2}-4)\frac{3c}{10}}
\approx -0.327962\,i\,,
\end{equation}
\noindent
which (by the analysis of eigenvalues of the Hessian matrix of
$-V_{\mathrm{bh}}$ with respect to $t$ and $t^{*}$, in a real basis
\cite{Bellucci:2006ew,Cardoso:2006cb}) is found to be
stable.\footnote{In each case there are in addition multiple
  stationary points outside of the allowed domain. For $q_{1}$ there
  is also one admissible saddle point of the black hole potential at
  $t = i[(3\sqrt{2}+4)\frac{3c}{10}]^{1/3} \approx 1.06216\,i$.}
Neither depends on the value of the charge.

The metric function of non-extremal solutions with the constant
scalar, fixed to one of the above values,
\begin{equation}
\label{eq:UUeconstt}
e^{-U} = e^{-r_{0}\tau}\left(\frac{-V_{\mathrm{bh}}\rvert_{\mathrm{att}}}{-2r_{0}^{2} \pm 2 \sqrt{r_{0}^{2}(r_{0}^{2}-V_{\mathrm{bh}}\rvert_{\mathrm{att}})}}(e^{2 r_{0}\tau}-1) + 1\right),
\end{equation}
\noindent
has the extremal ($r_{0} \to 0$) limit:
\begin{equation}
\lim_{r_{0} \to 0}e^{-U} = -\sqrt{-V_{\mathrm{bh}}}\Big\rvert_{\mathrm{att}}\tau + 1\,,
\end{equation}
\noindent
with the minus sign due to the negative $\tau$ in our conventions and
the constant $1$ for asymptotic flatness. The respective stationary
values of the black hole potential read
\begin{equation}
\label{eq:Vcrph1}
-V_{\mathrm{bh}}\Big\rvert_{\mathrm{att}} = -\frac{5 t_{\mathrm{att}}
  \left(144c^{2} + 30c t_{\mathrm{att}}^{3} + 100
    t_{\mathrm{att}}^{6}\right)}{8 \left(36c^{2} + 30c
    t_{\mathrm{att}}^{3} - 50
    t_{\mathrm{att}}^{6}\right)}(\hat{p}^{1})^2
\approx 2.20225(\hat{p}^{1})^{2}
\end{equation}
\noindent
and
\begin{equation}
-V_{\mathrm{bh}}\Big\rvert_{\mathrm{att}} =
\frac{\sqrt{2}}{2}\left(\frac{3\sqrt{2}+4}{75c}\right)^{1/3}(\hat{q}_{1})^{2}
\approx 0.431213(\hat{q}_{1})^{2}\,,
\end{equation}
the second of which does not have a finite $c\to 0$ limit.


\section{Conclusions}
\label{sec:conclusions}

The use of the H-FGK approach has enabled us, apart from studying some
model-independent properties of black-holes in four-dimensional $N=2$
supergravity, to find extremal and non-extremal solutions for the
$t^{3}$ model without and, for the first time analytically, with a
quadratic quantum correction to the prepotential. We study the
solutions for the corrected model in a symplectically rotated frame of
homogenous coordinates on the scalar manifold, which simplifies the
prepotential (and allows one to interpret the results as pairs of
solutions for two closely related, but not mutually dual prepotentials
with quadratic corrections).

The formalism itself can be applied with equal ease to any charge
configuration of either model, but the polynomial equations that
determine the parameters make the explicit solutions unfeasible except
when some charges vanish and, in the non-extremal case, when the
scalar is constant.

We find that the correction leads to the appearance of solutions,
which one might call quantum black holes, that do not possess a
regular classical limit. Perhaps surprisingly, we find in particular
that the quantum correction is sufficient to render the otherwise
divergent solution with only one charge, $q_{1}$, regular. (The other
solution that is single-charge in the rotated frame, but which is not
single-charge in the original frame, without the quantum correction
reduces to the empty Minkowski spacetime.)

In contrast to the solutions in ref.~\cite{Bueno:2012jc}, the
truncations ($H^{M} = 0$ for some $M$) of the $H$-functions
corresponding to our quantum black holes are non-singular in the
classical limit. This means that in our case we can construct the
classical counterpart to a corrected solution with no regular $c\to 0$
limit by simply considering the theory with $c=0$, imposing the same
constraints on $H^{M}$ and $\mathcal{Q}^M$, and then solving the
Freudhental duality equations.

\section*{Acknowledgments}

CSS would like to thank A.~Uranga for many useful conversations. JP
wishes to thank Prof.~G.L.~Cardoso for an e-mail discussion. This
work has been supported in part by the Spanish Ministry of Science and
Education grant FPA2009-07692, the Comunidad de Madrid grant HEPHACOS
S2009ESP-1473 and the Spanish Consolider-Ingenio 2010 program CPAN
CSD2007-00042. The work of PG has been supported in part by grants
FIS2008-06078-C03-02 and FIS2011-29813-C02-02 of Ministerio de Ciencia
e Innovaci\'on (Spain) and ACOMP/2010/213 from Generalitat
Valenciana. JP's work has been supported by the ERC Advanced Grant
no.~226455 (SUPERFIELDS). The work of CSS has been supported by a
JAE-predoc grant JAEPre 2010 00613. TO wishes to thank
M.M.~Fern\'andez for her permanent support.

\appendix

\section{Type~II Calabi--Yau compactifications}
\label{sec:CYC}

In this appendix we review the compactification of the type~IIA theory
on the quintic manifold $\mathcal{M}$ and of the type~IIB on the
mirror quintic manifold $\mathcal{W}$, following
refs.~\cite{Candelas:1985cyc,Witten:1985cyc,Candelas:1988cyc,Green:1989cyc,Plesser:1989cyc,Green:1990cyc,Candelas:1990cyc,Candelas:1991cyc,Ossa:1991cyc}. It
is well known that the low-energy limit of type~II superstring theory
compactified on a Calabi--Yau manifold is an $N=2,d=4$ supergravity
with a number of vector multiplets and hypermultiplets that depend on
the Hodge numbers of the Calabi--Yau manifold.  Only the vector
multiplets moduli space is relevant for the construction of black-hole
solutions in these theories: black-hole-type solutions with
non-trivial hyperscalars in ungauged $N=2,d=4$ theories are expected
to be generically singular since they would have primary scalar hair
\cite{Huebscher:2006mr}. On the other hand, in the unguaged theories,
the only bosonic field the hyperscalars couple to in the ungauged
theories is the metric, and, therefore, they can always be
consistently truncated or, equivalently, set to some constant value.


\subsection{Type~IIB on the mirror quintic $\mathcal{W}$}

Let $\mathcal{M}$ be the family of manifolds associated with the
vanishing of a quintic polynomial in $\mathbb{CP}_{4}$. An element of
$\mathcal{M}$ has $h^{(2,1)}=101$ degrees of freedom describing the
complex structure of the manifold, that can be associated with the
coefficients of the defining polynomial.\footnote{A quintic polynomial
  has 126 possible terms and complex coefficients. However, 25 of them
  can be eliminated by linear transformations of the 5 complex
  coordinates.} Furthermore, $h^{(1,1)}=1$ and the only independent
harmonic $(1,1)$-form can identified with the K\"ahler form of the
manifold: any other harmonic $(1,1)$-form is the K\"ahler form
multiplied by a real number, which corresponds to the freedom to
adjust the overall scale of the manifold. The Euler number of a
quintic manifold is $\chi =-200$.

Let us consider the family of quintic polynomials
\cite{Green:1989cyc,Plesser:1989cyc}
\begin{equation}
\label{eq:mirrorpolynomial}
p_{\psi}=\sum^{5}_{k=1} x^{5}_{k} - 5\psi\prod^{5}_{k=1} x_k,\qquad
\psi\in\mathbb{C}\, ,
\end{equation}
\noindent
parametrized by the complex modulus $\psi$, $M_{\psi}$ the manifold
described by $p_{\psi}=0$ and $\mathcal{M}_{0}\subset \mathcal{M}$ the
family of all manifolds $M_{\psi}$ for $\psi\in \mathbb{C}$ The family
of quintic polynomials $\left( p_{\psi},~\psi\in\mathbb{C} \right)$ is
invariant under the group generated by:
\begin{equation}
 \begin{array}{rcl}
g_{0} & = & \left( 1, 0,0,0,4\right)\, ,\\
g_{1} & = & \left( 0, 1,0,0,4\right) \, ,\\
g_{2} & = & \left( 0, 0,1,0,4\right)\, ,\\
g_{3} & = & \left( 0, 0,0,1,4\right)\, ,\\
 \end{array}
\end{equation}
\noindent
where $g_{i}$, $i=0,\dotsc,3$, acts on $\left(x_{1},\dotsc,x_{5}\right)$
by multiplying the $\left(i+1\right)$-th entry by the phase $\alpha=e^{2\pi
  i/5}$ and the last entry by $\alpha^{4}$, so $g_{i}^{5}=1$ for all $i$. The
transformation $g_{0}g_{1}g_{2}g_{3}$ leaves each $p_{\psi}$ invariant because
it multiplies the homogeneous coordinates by a common phase, hence only
three of the $g_{i}$ are independent, say $g_{1},g_{2}$ and $g_{3}$.  These
three elements generate the group $\mathbb{Z}^{3}_{5}$.

It turns out that the mirror family $\mathcal{W}$ is $\mathcal{W}= W_{\psi}
\equiv M_{\psi}/\mathbb{Z}^{3}_{5}, \psi \in \mathbb{C}$. It can be
shown that the elements of $\mathcal{W}$ have $h^{(2,1)} = 1$, $h^{(1,1)}=101$
and $\chi = 200$, as they must.

Since the transformation $\psi\rightarrow \alpha\psi$ can be undone by a
coordinate transformation, we have that $\psi\sim\alpha\psi$, thus it is
$\psi^{5}$ that plays the role of the modulus that parametrizes the
complex-structure moduli space of $\mathcal{W}$ that we denote by
$C^{(2,1)}_{\rm IIB}$. This is in agreement with $h^{(2,1)}=1$. There are two
values of $\psi^{5}$ for which $M_{\psi}$ (and, correspondingly, $W_{\psi}$)
is singular: $\psi^{5}=1$ and $\psi=\infty$.

$W_{1}$ has a single singular point given by the equivalence class
$\left[\left(1,1,1,1\right)\right]$ and $W_{\infty}$ is given by the quotient
by $\mathbb{Z}^{3}_{5}$ of the singular quintic
\begin{equation}
\label{eq:mirrorpolynomialinfty}
p_{\infty}=\prod^{5}_{k=1} x_k=0\, .
\end{equation}

$W_{\infty}$ is the large complex structure limit of $\mathcal{W}$: we will
see in the following section that it is the mirror of the large-radius limit
of $\mathcal{M}$.

We are interested in the compactification of the type~IIB theory on
$\mathcal{W}$. The low-energy effective field theory is an ungauged $N=2,d=4$
supergravity coupled to $h^{(2,1)}=1$ vector multiplets and $h^{(1,1)}+1=102$
hypermultiplets that can be consistently ignored (set to some constant
value). We will thus be dealing with just one complex scalar parametrizing
the special K\"ahler manifold $C^{(2,1)}_{\rm IIB}$.

Following ref.~\cite{Candelas:1991cyc}, we can describe the
complex-structure moduli space $C^{(2,1)}_{\rm IIB}$ by the periods of
the holomorphic three-form $\Omega$ over a canonical basis of
$H_{3}( W_{\psi},\mathbb{Z})$, which in our case, since
$b_{3}=4$, can be taken to be $(\gamma^{M}) = \left(A^{0}, A^{1}, B_{0},
  B_{1}\right)^{\mathrm{T}}$ with the intersections
\begin{equation}
\label{eq:homologybasis}
A^{\Lambda}\cap B_{\Gamma} = \delta^{\Lambda}_{\Gamma}\, ,
\qquad
A^{\Lambda}\cap A^{\Gamma}  = 0\, ,
\qquad
B_{\Lambda} \cap B_{\Gamma} =0\, .
\end{equation}
\noindent
The dual cohomology basis is denoted by $\left(\alpha_{\Lambda},\beta^{\Gamma}
\right)$ and obeys
\begin{equation}
\label{eq:dualbasiscondition}
\int_{A^{\Lambda}}\alpha_{\Gamma}  = \delta^{\Lambda}{}_{\Gamma}\, ,
\qquad
\int_{B_{\Lambda}}\beta^{\Gamma}  = -\delta^{\Gamma}{}_{\Lambda}\, ,
\qquad
\int_{A^{\Lambda}}\beta^{\Gamma}  =
\int_{B_{\Lambda}}\alpha_{\Gamma}  = 0\, .
\end{equation}
\noindent
The holomorphic 3-form $\Omega$ is given by
\begin{equation}
\label{eq:Omega}
\Omega = \mathcal{X}^{\Lambda} \alpha_{\Lambda}
-\mathcal{F}_{{\rm IIB},\Lambda} \beta^{\Lambda}\, ,
\end{equation}
\noindent
where $\mathcal{X}^{\Lambda}$ and $\mathcal{F}_{{\rm IIB}\, \Lambda}$, which
will be identified as the components of the holomorphic symplectic section
\begin{equation}
\Pi_{\rm IIB}(\psi) =
\left(
 \begin{array}{c}
\mathcal{X}^{0} \\
\mathcal{X}^{1}\\
\mathcal{F}_{{\rm IIB}\,0}\\
\mathcal{F}_{{\rm IIB}\,1}\\
 \end{array}
\right) ,
\end{equation}
\noindent
are the periods of the holomorphic 3-form with respect to the canonical
homology basis
\begin{equation}
\label{eq:periods}
\mathcal{X}^{\Lambda} = \int_{A^{\Lambda}} \Omega\, ,
\qquad
\mathcal{F}_{\Lambda} = \int_{B_{\Lambda}}\Omega\, .
\end{equation}

There are 4 periods, but the complex-structure manifold is one-dimensional and
hence we can take the $\mathcal{F}_{\Lambda}$ to be holomorphic functions of
the $\mathcal{X}^{\Lambda}$. Since $\Omega$ is defined up to rescalings
$\Omega\rightarrow g(\psi)\Omega$, where $g(\psi)$ is a
holomorphic function of the modulus $\psi$, we can take the
$\mathcal{X}^{\Lambda}$ to be projective coordinates of the scalar manifold,
and hence we end up with one complex coordinate, which is what we need in
order to parametrize $C^{(2,1)}_{\rm IIB}$. Different choices of $g(\psi)$ can
be understood as different gauge choices. In addition, the periods
$\mathcal{F}_{{\rm IIB}\, \Lambda}$ can be expressed as derivatives of a
single function $\mathcal{F}_{\rm IIB}$ of the $\mathcal{X}^{\Lambda}$:
\begin{equation}
\mathcal{F}_{{\rm IIB}\, \Lambda}
=
\frac{\partial \mathcal{F}_{\rm IIB}}{\partial\mathcal{X}^{\Lambda}}\, .
\end{equation}

We will find later on that it is more natural to consider
$\mathcal{F}_{{\rm IIB}\, \Lambda}$ as the projective coordinates and the
$\mathcal{X}^{\Lambda}$ given in terms of them. A good special
coordinate in the large complex-structure limit is therefore provided
by:
\begin{equation}
\label{eq:specialcoordinate}
Z(\psi)
=
\frac{\mathcal{F}_{{\rm IIB}\, 0}(\psi)}{\mathcal{F}_{{\rm IIB}\,
    1}(\psi)}\, .
\end{equation}

It can be shown \cite{Ferrara:1991aj,Ceresole:1992su} that the components of
the holomorphic symplectic section of an $N=2,d=4$ supergravity theory have to
obey a set of differential identities due to the properties of the
special K\"ahler geometry. When the theory originates from a Calabi--Yau
compactification, these identities are the Picard--Fuchs equations. In our
case, there is only one fourth-order Picard--Fuchs equation associated with
$\mathcal{W}$ \cite{Ferrara:1991aj,Klemm:1992tx}
\begin{equation}
\label{eq:PFeq}
(1-\psi^{5})\omega^{iv} - 10 \psi^{4} \omega^{\prime\prime\prime}
- 25 \psi^{3}\omega^{\prime\prime}-15\psi^{2} \omega^{\prime} - \psi\, \omega =0\, .
\end{equation}
\noindent
and its 4 independent solutions $\omega_{0},\omega_{1},\omega_{2},\omega_{3}$
can be identified with the 4 periods \cite{Ossa:1991cyc}.

Eq.~(\ref{eq:PFeq}) is an ordinary differential equation with regular singular
points at $\psi^{5}=0,1,\infty$ and, hence, a system of solutions may be
obtained following the method of Froebenius for such equations. At $\psi^{5}
=\infty$ one solution, $\omega_{0}$, is given as a pure power series and the
other three solutions $\omega_{1},\omega_{2},\omega_{3}$ contain logarithms,
with powers 1, 2 and 3, respectively. At $\psi^{5}=0$ all four solutions are
pure power series. We will not need the solutions at $\psi^{5}=1$.

The pure power series solution around $\psi^{5} =\infty$ is
\begin{equation}
\label{eq:omega0+}
\omega_{0}(\psi) =
\frac{1}{5\psi}\sum^{\infty}_{n=0}
\frac{\left(5n\right)!}{\left(n!\right)^{5}\left(5\psi\right)^{5n}}\, ,
\qquad
|\psi|>1\, ,
\qquad
0\le \operatorname{Arg}(\psi) < \frac{2\pi}{5}\, .
\end{equation}

This expression has been obtained with the choice of $g(\psi)$ normally used
to study the (mirror) Landau--Ginzburg or Fermat limit $\psi\rightarrow 0$.
An  expression for $\omega_{0}$ in the large complex structure limit can be obtained from the one above by a gauge transformation with $g(\psi)=5\psi$ \cite{Klemm:1992tx} that gets
rid of the overall factor $\left(5\psi\right)^{-1}$. We will use this new
gauge for both limits, since we have found no complications in using it in the
Fermat limit $\psi\rightarrow 0$. In conclusion, we take
$\omega_{0}$ to be
\begin{equation}
\label{eq:omega0+ginf}
\omega_{0}(\psi)
=
\sum^{\infty}_{n=0}
\frac{\left(5n\right)!}{\left(n!\right)^{5}\left(5\psi\right)^{5n}}\, ,
\qquad
|\psi|>1\ ,
\qquad
0\le \operatorname{Arg}(\psi) < \frac{2\pi}{5}\, .
\end{equation}

The solution around $\psi=0$ can be obtained by analytical continuation of
eq.~(\ref{eq:omega0+ginf}):
\begin{equation}
\label{eq:omega0-}
\omega_{0}(\psi) =
-\frac{1}{5}\sum^{\infty}_{m=1}
\frac{\alpha^{2m}\Gamma\left(m/5\right)\left(5\psi\right)^m}
{\Gamma\left(m\right)\Gamma^{4}\left(1-m/4\right)}\, ,
\qquad|\psi|<1\, .
\end{equation}

The 5 functions
\begin{equation}
\omega_{k}(\psi)\equiv \omega_{0}(\alpha^{k}\psi)\, ,\qquad
 k=0,\dotsc,4\, ,
\end{equation}
\noindent
are also solutions, but one of them cannot be linearly independent: 
the $\omega_{k}$ obey a linear relation which turns out to be
\begin{equation}
\sum^{4}_{k=0}\omega_{k}=0\, .
\end{equation}
\noindent
The expressions for the $\omega_{k}$, $k=1,\dotsc,4$ for $|\psi|>1,
0<\operatorname{Arg}(\psi)<\frac{2\pi}{5}$ are quite involved and can be found in
appendix~\ref{sec:largecompexstructurelimit}.

To construct the holomorphic symplectic section $\Pi_{\rm IIB}$ we
choose a set of four linearly independent solutions, that we combine  into a
vector $\hat{\omega}$ (also called the period vector on the Picard--Fuchs
basis)
\begin{equation}
\hat{\omega} =
-\left(\frac{2\pi i}{5}\right)^{3}
\left(
 \begin{array}{c}
\omega_{2} \\
\omega_{1}\\
\omega_{0}\\
\omega_{4}\\
 \end{array}
\right)\, ,
\end{equation}
\noindent
and then define $\Pi_{IIB}(\psi)$ by
\begin{equation}
\label{eq:PIIIB}
\Pi_{\rm IIB}(\psi)
= M\, \hat{\omega}
\qquad
M=\left( \begin{array}{cccc}
-1 & 0 & 8 & 3\\
0 & 1 & -1 & 0\\
-3/5 & -1/5 & 21/5 & 8/5 \\
0 & 0 & -1 & 0\\
\end{array} \right)\, .
\end{equation}
\noindent
The K\"ahler potential is given by
\begin{equation}
\label{eq:KIIB}
e^{-\mathcal{K}}
=
i \left(\mathcal{X}^{*\Sigma}\mathcal{F}_{{\rm IIB}\, \Sigma}
-\mathcal{X}^{\Sigma}\mathcal{F}^{*}_{{\rm IIB}\, \Sigma} \right)
= \omega^{\dagger}\sigma\, \omega\, ,
\end{equation}
\noindent
where
\begin{equation}
\sigma \equiv
\frac{1}{5}
\left(
\begin{array}{cccc}
0 & 1 & 3 & 1 \\
-1 & 0 & 3 & 3\\
-3 & -3 & 0 & 1\\
-1 & -3 & -1 & 0\\
\end{array}
\right)\, .
\end{equation}
\noindent
Eq.~(\ref{eq:KIIB}) is a very complicated function of $\psi$, hence
some simplification limit is in order. It can be shown that in the
large complex-structure limit (given by eq.~(\ref{omegaaprox}))
$\psi\rightarrow \infty$ the K\"ahler potential is given by:
\begin{equation}
\label{eq:KIIBaprox}
e^{-\mathcal{K}}
=
\left(\frac{2\pi}{5}\right)^{3}\left(\frac{20}{3} \log^{3}|5\psi|
+\frac{16}{5}\zeta(3)\right)\, .
\end{equation}
\noindent
From (\ref{eq:KIIBaprox}) we can compute the K\"ahler metric
\begin{equation}
\label{eq:gIIBaprox1}
\mathcal{G}_{\psi\psi^{*}}
=
\frac{15 \left(-24 \zeta \left(3\right) \log|5\psi|
+5 \log^{4} | 5\psi |\right)}{|\psi|^{2} \left( 24 \zeta\left( 3\right)
+10 \log^{3} |5\psi|\right)}\, .
\end{equation}
\noindent
We can expand (\ref{eq:gIIBaprox1}) as to obtain:
\begin{equation}
\label{eq:gIIBaprox2}
\mathcal{G}_{\psi\psi^{*}}
=
\frac{3}{4|\psi|^{2} \log^{2} |5\psi|}\left(1-\frac{48\zeta(3)}{25
    \log^{3} |5\psi|}
+\dotsb \right) .
\end{equation}
\noindent
We perform the change of variable
\begin{equation}
t\equiv -\frac{5}{2\pi i} \log(5\psi)
\end{equation}
\noindent
in order to make easier the comparison with the metric of the large-radius
limit of type~IIA on $\mathcal{M}$, which is obtained in the following
section. The leading term of (\ref{eq:gIIBaprox2}) becomes
\begin{equation}
\label{eq:gtIIBaprox}
\mathcal{G}_{tt^{*}} = \tfrac{3}{4}(\Im\mathfrak{m}\, t)^{-2}\, ,
\end{equation}
\noindent
which is, as we will see, the large-radius limit metric of the
K\"ahler-structure moduli space, the scalar manifold of type~IIA on
$\mathcal{M}$.

\subsection{Type~IIA on the quintic $\mathcal{M}$ and mirror map}
\label{sec:TypeIIA}

The low-energy effective theory of of type~IIA superstring theory compactified
on a Calabi--Yau manifold is $N=2$ supergravity coupled to $h^{(1,1)}$ vector
multiplets and $h^{(2,1)}+1$ hypermultiplets. The prepotential in the large
compactification radius limit is given by \cite{Candelas:1991cyc}
\begin{equation}
\label{eq:preptreelevel}
\mathcal{F}^{0}_{\rm IIA}(\mathcal{X})
=
-\frac{1}{3!}\frac{\kappa^{0}_{ijk}
\mathcal{X}^{i}\mathcal{X}^{j}\mathcal{X}^{k}}{\mathcal{X}^{0}}\, ,
\qquad
i,j,k=1,\dotsc,h^{(1,1)}\, .
\end{equation}
\noindent
where $\kappa^{0}_{ijk}$ are the triple intersection numbers.

We take the compactification manifold to be quintic $\mathcal{M}$,
hence $h^{(1,1)}=1$ and $h^{(2,1)}=101$. Since, as in the type~IIB
case, we are only interested in the vector multiplet moduli space, we
set the hypermultiplets to zero and deal solely with the complex
K\"ahler-structure moduli space $C^{(1,1)}_{\rm IIA}$, which is a
complex one-dimensional special K\"ahler manifold.

If we denote by $e$ the generator of $H^{2}(\mathcal{M},
  \mathbb{Z})$, the only non-vanishing triple intersection number at tree
level is
\begin{equation}
\kappa^{0}_{1,1,1}=\int_{\mathcal{M}} e\wedge e\wedge e = 5\, .
\end{equation}
\noindent
Then, in terms of the coordinate
\begin{equation}
\label{eq:tdef}
t \equiv \mathcal{X}^{1}/\mathcal{X}^{0}
\end{equation}
\noindent
and in the K\"ahler gauge $\mathcal{X}^{0}=1$, the K\"ahler potential
is given by
\begin{equation}
\label{eq:KIIAtreelevel}
\mathcal{K}_{\rm IIA}^{0}
=
-\log\left[\tfrac{20}{3} (\Im \mathfrak{m}\, t)^{3}\right].
\end{equation}
\noindent
The K\"ahler metric reads
\begin{equation}
\label{eq:gIIAtreelevel}
\mathcal{G}^{0}_{tt^{*}}
=
\tfrac{3}{4}\left(\Im\mathfrak{m}\, t\right)^{-2}\, .
\end{equation}
\noindent
Comparing eqs.~(\ref{eq:gtIIBaprox}) and~(\ref{eq:gIIAtreelevel}) we can see
that the large complex-structure limit of the metric of $C^{(2,1)}_{\rm IIB}$
agrees with the corresponding bare (uncorrected) quantities for
$C^{(1,1)}_{\rm IIA}$.

We are interested in how the loop corrections and worldsheet instanton
corrections (we restrict ourselves to a two-derivative action) to
eq.~(\ref{eq:preptreelevel}) affect non-extremal black-hole
solutions. One can write the corrected prepotential
\cite{Ossa:1991cyc} in the form $\mathcal{F}_{\rm
  IIA}=\mathcal{F}^{\rm pert}_{\rm IIA}+\mathcal{F}^{\rm npert}_{\rm
  IIA}$, where $\mathcal{F}^{\rm pert}_{\rm IIA}$ denotes the
perturbatively-corrected prepotential and $\mathcal{F}^{\rm
  npert}_{\rm IIA}$ denotes the exponentially small terms due to
instanton corrections. They are given by:
%
%
\begin{eqnarray}
\label{eq:FIIAPert}
\mathcal{F}^{\rm pert}_{\rm IIA}
& = &
\mathcal{F}^{0}_{\rm IIA}+\mathcal{F}^{\rm loop}_{\rm IIA}
=
-\frac{5}{6} \frac{\left(\mathcal{X}^{1}\right)^{3}}{\mathcal{X}^{0}}
-\frac{11}{4} (\mathcal{X}^{1})^{2} +\frac{25}{12} \mathcal{X}^{0}\mathcal{X}^{1}
-ik (\mathcal{X}^{0})^{2}, \\
& & \nonumber \\
\label{eq:FIIANonPert}
\mathcal{F}^{\rm npert}_{\rm IIA}
& = &
\sum_{l} n_{l} \operatorname{Li}_{3} \left( e^{2\pi i l \mathcal{X}^{1}/\mathcal{X}^{0}}\right) ,
\end{eqnarray}
\noindent
where
\begin{equation}
\operatorname{Li}_{3}(x)=\sum^{\infty}_{j=1}\frac{x^{j}}{j^{3}}\, ,
\end{equation}
\noindent
and $n_{k}$ is the number of rational curves of degree $k$, and where we have
defined the real numerical constant
\begin{equation}
\label{eq:cdef}
k \equiv \frac{c}{2} \equiv \frac{25}{2\pi^{3}}\zeta(3)\, .
\end{equation}
\noindent
For large values of the quintic radius $\Im \mathfrak{m}\, t \gg 1$, the
non-perturbative contribution to the prepotential are exponentially small and
can be ignored.

The type~IIB theory compactified on $\mathcal{W}$ is related to the
type~IIA one compactified on $\mathcal{M}$ through the mirror map,
which can be expressed as a symplectic transformation of the
holomorphic symplectic section with matrix $N$ given by
\cite{Ossa:1991cyc}
\begin{equation}
\Pi_{\rm IIA}
=
\frac{\mathcal{F}_{{\rm IIB}\, 1}}{\mathcal{X}^{0}} N \Pi_{\rm IIB}\, ,
\qquad
N
=
\left(
\begin{array}{cccc}
0 & 0 & 0 & 1 \\
-1 & 0 & 2 & 0\\
0 & -1 & 0 & 0\\
0 & 0 & -1 & 0
\end{array}
\right)\, ,
\end{equation}
\noindent
and the coordinate transformation
\begin{equation}
\label{eq:MirrorMap}
t=\frac{2\left(\omega_{1}-\omega_{0}\right)+\omega_{2}-\omega_{4}}{5\omega_{0}}\, ,
\end{equation}
\noindent
where we are denoting the holomorphic symplectic section of the type~IIA
theory compactified on $\mathcal{W}$ by
\begin{equation}
\label{eq:periodIIA}
\Pi_{\rm IIA}(\psi)
=
\left(
\begin{array}{c}
\mathcal{X}^{0} \\
\mathcal{X}^{1} \\
\mathcal{F}_{{\rm IIA}\, 0}\\
\mathcal{F}_{{\rm IIA}\, 1} \\
\end{array}
\right)\, .
\end{equation}
\noindent
Consequently, at the supergravity level, both theories are the same theory in
different coordinates and symplectic frames.


\subsection{Large complex-structure limit}
\label{sec:largecompexstructurelimit}

In this section we give the explicit expressions for the periods in
region $|\psi|>1, 0 \le \operatorname{Arg}(\psi) < \frac{2\pi}{5}$, and
we also obtain the large complex-structure limit
\cite{Ossa:1991cyc}. The periods are given by:
\begin{equation}
\omega_j(\psi) = \sum^{3}_{r=0}
\log^r(5\psi)\sum^{\infty}_{n=0} b_{jrn}
\frac{(5\psi)!}{(n!)^{5}(5\psi)^{5n}}\,,
\qquad |\psi|>1\, ,
\label{eq:omega0infresto}
\end{equation}
\noindent
where the coefficients are given by lengthy expressions that can be
found in \cite{Ossa:1991cyc}. In the large complex-structure limit
$\psi\to\infty$ we keep the first term in the pure power expansion
of eq.~(\ref{eq:omega0infresto}). We can then write a vector of
coefficients:
\begin{equation}
b_r =-\left(\frac{2\pi i}{5}\right)^{3}\left(
 \begin{array}{rcl}
b_{2r0}\\
b_{1r0}\\
b_{0r0}\\
b_{4r0}
 \end{array}
\right) ,
\label{br}
\end{equation}
\noindent
in terms of which the large complex-structure limit of the period
vector in Picard--Fuchs basis is written as:
\begin{equation}
\hat{\omega}\sim \sum^{3}_{r=0}b_r \log^r(5\psi)\, .
\label{omegaaprox}
\end{equation}
\noindent
Eq.~(\ref{omegaaprox}) is the starting point for obtaining the
relevant quantities of the model in the limit $\psi\to\infty$.


\subsection{A simpler prepotential}
\label{sec:AsimplerPrep}

As already mentioned, for large values of the quintic radius $\Im
\mathfrak{m}\, t\gg 1$, the non-perturbative contributions to the
prepotential are exponentially small, so $\mathcal{F}^{\rm npert}_{\rm
  IIA}$ of eq.~(\ref{eq:FIIANonPert}) can be neglected. Taking into
account just eq.~(\ref{eq:FIIAPert}), the holomorphic symplectic
section is given by
\begin{equation}
\label{eq:correctedsection}
\Pi_{\rm pert}
=
\left(
\begin{array}{c}
\mathcal{X}^{0} \\
\\
\mathcal{X}^{1} \\
\\
{\displaystyle
\frac{5}{6}\frac{(\mathcal{X}^{1})^{3}}{(\mathcal{X}^{0})^{2}}
+\frac{25}{12}\mathcal{X}^{1} -i c \mathcal{X}^{0}
}
\\
\\
{\displaystyle
-\frac{5}{2}\frac{(\mathcal{X}^{1})^{2}}{\mathcal{X}^{0}}
-\frac{11}{2}\mathcal{X}^{1}
+\frac{25}{12}\mathcal{X}^{0}
}
\\
\end{array}
\right)\, ,
\end{equation}
\noindent
In the spirit of ref.~\cite{Bellucci:2010zd}, the symplectic
(Peccei--Quinn) transformation
\begin{equation}
\hat{\mathcal{S}}
\equiv
\left(
\begin{array}{c|c}
\mathbb{I} & 0 \\
\hline
&  \\
\begin{array}{cc}
0 &- \frac{25}{12} \\
& \\
-\frac{25}{12}& \frac{11}{2} \\
\end{array}
&
\mathbb{I} \\
\end{array}
\right)\, ,
\end{equation}
\noindent
brings the section to the simpler form
\begin{equation}
\label{eq:PiPert}
\hat{\Pi}_{\rm pert}
=
\left(
\begin{array}{c}
 \hat{ \mathcal{X}}^{0} \\
\\
\hat{\mathcal{X}}^{1} \\
\\
{\displaystyle\frac{5}{6}\frac{(\hat{\mathcal{X}}^{1})^{3}}{(\hat{\mathcal{X}}^{0})^{2}}}
-i c \hat{\mathcal{X}}^{0}
\\
\\
{\displaystyle-\frac{5}{2}\frac{(\hat{\mathcal{X}}^{1})^{2}}{\hat{\mathcal{X}}^{0}}}
\\
\end{array}
\right)\, ,
\end{equation}
\noindent
which can be derived from the prepotential
\begin{equation}
\label{eq:FIIAPertnew}
\hat{\mathcal{F}}^{\rm pert}_{\rm quintic}
=
-\frac{5}{6} \frac{(\hat{\mathcal{X}}^{1})^{3}}{\hat{\mathcal{X}}^{0}}
-\frac{i}{2} c (\hat{\mathcal{X}}^{0})^{2}\, .
\end{equation}
\noindent

The geometry of the scalar manifold in the corrected case is quite
different from $SL(2,\mathbb{R})/U(1)$ of the pure $t^3$ model. It is
not a homogeneous space and the conditions that $\Im\mathfrak{m}\, t$
has to satisfy are also different: the K\"ahler potential is given by
\begin{equation}
\label{eq:Kcorrected}
e^{-\mathcal{K}_{\rm pert}}
=
\tfrac{20}{3} (\Im\mathfrak{m}\, t)^{3} +2 c
\end{equation}
\noindent
and the fact that $\mathcal{K}$ must be real implies
\begin{equation}
\label{eq:tconstraint1}
\Im\mathfrak{m}\, t > -\left(\tfrac{3}{10} c\right)^{1/3}\, .
\end{equation}
\noindent
The K\"ahler metric is given by
\begin{equation}
\label{eq:Kmetriccorrected}
\mathcal{G}_{t\bar{t}}
=
\frac{15\, \Im\mathfrak{m}\, t \left[-3 c + 5 (\Im\mathfrak{m}\,
    t)^{3}\right]}{\left[3 c + 10 (\Im\mathfrak{m}\, t)^{3}\right]^{2}}\, .
\end{equation}
\noindent
For it to be positive definite, we need to demand $\Im\mathfrak{m}\,
t\, \left[-3 \epsilon c + 5 (\Im\mathfrak{m}\, t)^{3}\right]>0$. This
condition, together with eq.~(\ref{eq:tconstraint1}), gives the domain
of definition for $\Im\mathfrak{m}\, t$:
\begin{equation}
\label{eq:tconstraint2}
\Im\mathfrak{m}\, t\in \left( -\left(\frac{3 c}{10}\right)^{1/3},0\right)
\cup\left( \left(\frac{3 c}{5}\right)^{1/3},\infty\right)\, .
\end{equation}
\noindent
From the point of view of the supergravity theory, this is the only
condition that the scalar needs to satisfy for the solution to be well
defined. If, however, this supergravity is to be seen as an effective
description of the underlying superstring theory, there are more
conditions to be met by $t$. In particular, the prepotential
(\ref{eq:FIIAPertnew}) is an expansion around $t\to \infty$, valid
only inside the radius of convergence:
\begin{equation}
\label{eq:tconstraint3}
\Im\mathfrak{m}\, t > \Im\mathfrak{m}\, t(1)\, ,
\end{equation}
\noindent
where $t(\psi)$ is the mirror map, $\psi$ is the modulus of the mirror
related theory, and the conifold point is assumed to be at $\psi=1$.


\bibliographystyle{JHEP}
\bibliography{HFGKD4Q}

\end{document}